\documentclass[fleqn,usenatbib]{mnras}

\usepackage{newtxtext,newtxmath}

\usepackage[T1]{fontenc}

\DeclareRobustCommand{\VAN}[3]{#2}
\let\VANthebibliography\thebibliography
\def\thebibliography{\DeclareRobustCommand{\VAN}[3]{##3}\VANthebibliography}

\usepackage{graphicx}	
\usepackage{amsmath}	
\usepackage{subfig}
\usepackage{enumerate}
\usepackage{enumitem}

\title[Time varying Na\,\textsc{i} D absorption in ILRTs]{Time varying Na\,\textsc{i} D absorption in ILRTs as a probe of circumstellar material}

\author[R. A. Byrne et al.]{
R. A. Byrne,$^{1}$\thanks{E-mail: robert.byrne.6@ucdconnect.ie (RB)}
M. Fraser,$^{1}$
Y.-Z. Cai,$^{2,3}$
A. Reguitti,$^{4,5,6}$
G. Valerin$^{6,7}$
\\
$^{1}$School of Physics, University College Dublin, Belfield, Dublin 4, Ireland\\
$^{2}$Yunnan Observatories, Chinese Academy of Sciences, Kunming 650216, PR China\\
$^{3}$Key Laboratory for the Structure and Evolution of Celestial Objects, Chinese Academy of Sciences, Kunming 650216, PR China\\
$^{4}$Instituto de Astrofìsica, Departamento de Ciencias Fisicas, Universidad Andres Bello, Fernandez Concha 700, Las Condes, Santiago 8320000, Chile\\
$^{5}$Millennium Institute of Astrophysics (MAS), Nuncio Monsenor Sòtero Sanz 100, Providencia, Santiago 8320000, Chile\\
$^{6}$INAF — Osservatorio Astronomico di Padova, Vicolo dell’Osservatorio 5, 35122 Padova, Italy\\
$^{7}$Dipartimento di Fisica e Astronomia “G. Galilei”, Università degli studi di Padova Vicolo dell’Osservatorio 3, I-35122 Padova, Italy
}

\date{Accepted XXX. Received YYY; in original form ZZZ}

\pubyear{2023}

\begin{document}
\label{firstpage}
\pagerange{\pageref{firstpage}--\pageref{lastpage}}
\maketitle

\begin{abstract}
Intermediate-Luminosity Red Transients (ILRTs) are a class of observed transient posited to arise from the production of an electron-capture supernova from a super-asymptotic giant branch star within a dusty cocoon. In this paper, we present a systematic analysis of narrow Na\,\textsc{i} D absorption as a means of probing the circumstellar environment of these events. We find a wide diversity of evolution in ILRTs in terms of line strength, time-scale, and shape. We present a simple toy model designed to predict this evolution as arising from ejecta from a central supernova passing through a circumstellar environment wherein Na\,\textsc{ii} is recombining to Na\,\textsc{i} over time. We find that while our toy model can qualitatively explain the evolution of a number of ILRTs, the majority of our sample undergoes evolution more complex than predicted. The success of using the Na\,\textsc{i} D doublet as a diagnostic tool for studying circumstellar material will rely on the availability of regular high-resolution spectral observations of multiple ILRTs, and more detailed spectral modelling will be required to produce models capable of explaining the diverse range of behaviours exhibited by ILRTs. In addition, the strength of the Na\,\textsc{i} D absorption feature has been used as a means of estimating the extinction of sources, and we suggest that the variability visible in ILRTs would prevent such methods from being used for this class of transient, and any others showing evidence of variability.
\end{abstract}





\section{Introduction} \label{sec:intro}

With the advent of modern astronomical surveys, the `luminosity gap' \citep[e.g.][]{kulkarnigap, pastorellogap, caigap}, spanning from the brightest classical novae to the dimmest core-collapse supernovae has begun to be populated with a number of new classes of intermediate-luminosity transients. Among these are the intermediate-luminosity red transients (ILRTs).

In terms of photometry, ILRTs typically show a slow rise of $\sim$ 2 weeks to a peak absolute magnitude between $-11.5$ and $-14.5$ mag in the \textit{V}-band. A linear decline or pseudo-plateau phase follows the peak, leading the shape of their light curves to resemble those of Type IIL or Type IIP supernovae respectively. In cases where late-time photometry has been available, their decline has been seen to match the decay of $^{56}$Co. With total radiated energies on the order of $10^{47}$ erg, they are fainter than the majority of Type II supernovae.

Spectra from ILRTs tend to be relatively featureless, with narrow H~$\alpha$ and H~$\beta$ emission lines being the most prominent features, alongside the [Ca\,\textsc{ii}] doublet ($\lambda\lambda$ 7291, 7324) and the Ca infrared triplet ($\lambda\lambda$ 8498, 8542, 8662). Lines such as Na\,\textsc{i} D and Ca H\&K can also be seen in absorption. The [Ca\,\textsc{ii}] emission doublet ($\lambda\lambda$ 7291, 7324) is strongly visible throughout the duration of the transient and is considered a characteristic feature of ILRTs. The spectra evolve slowly, becoming slightly redder over time, and exhibiting an IR excess at both early and late times, suggestive of a dusty local environment.

In addition to the prototypical event for this class: SN 2008S \citep{botticella08s}, a handful of these transients have been discovered and studied such as NGC 300-2008OT1 \citep{bond300ot, berger, adams300ot}, AT 2017be \citep{cai17be}, and AT 2019abn \citep{jencson19abn}. A spectroscopic and photometric study of five ILRTs \citep{caiILRTs} shows strong homogeneity between members of this class of transient.

A number of mechanisms have been suggested for how the transient itself is produced. Some possibilities include outbursts similar to those of Luminous Blue Variables (LBVs) \citep{smithLBVs, humphreys300ot}, a stellar merger similar to a Luminous Red Nova (LRN) \citep{kasliwalLRNe}, or a faint core-collapse supernova produced by the core of a super-asymptotic giant branch star undergoing electron-capture \citep{botticella08s, dohertySAGB}.

Progenitor candidates have been detected for a number of ILRTs through pre-explosion imaging. MIR imaging from \textit{Spitzer} revealed a source coincident with the position of SN 2008S consistent with a $\sim 10$ M$_\odot$ star enshrouded in a cloud of dust at a temperature of $\sim 440$ K \citep{prieto08sprogen}. Similar archival observations at the position of NGC 300-2008OT1 prior to explosion show a possible progenitor with a luminosity of $10^{4.9}$ L$_\odot$ and a dust temperature of $\sim$ 300 K \citep{adams300ot}. \cite{jencson19abn} identify a similar progenitor for AT 2019abn, which exhibits variability in the 4.5 micron band in the years prior to explosion.

Follow-up campaigns for SN 2008S and NGC 300-2008OT1 have confirmed that both objects are still fading, and are > 15 times fainter than their progenitors in the MIR, and undetected in optical and NIR \citep{adams300ot}. This lends credence to those models which imply these events arise from terminal explosions rather than non-terminal eruptions.

In a number of ILRTs, the strength of the Na\,\textsc{i} D ($\lambda\lambda$ 5890, 5896) absorption doublet has been seen to vary with time \citep{botticella08s, caiILRTs}. Such variability has also been measured in other classes of gap transient, such as the luminous red nova AT 2021biy \citep{21biy}. This variability may suggest that these lines are being produced in clouds of circumstellar material around the progenitor, and by tracking their evolution, we may be able to probe these dusty environments. In doing so, we can use insights from the circumstellar environment to infer the type of progenitors which may produce such environments.

The Na\,\textsc{i} D line has been used to probe a number of classes of SNe. These include the Type Ia supernova SN 2006X, where the evolution of this feature was interpreted as occurring due to changes in the ionisation conditions of the circumstellar material (CSM) due to radiation from the supernova \citep{patat06x}. Na\,\textsc{i} D absorption has also been studied for large statistical samples of Type Ia SNe, where a preponderance of blue-shifted absorption has been taken as evidence for outflowing gas from their progenitor systems \citep{Sternberg2011,Maguire2013}. Detections of Na\,\textsc{i} D absorption from the Type IIL supernova SN 1998S have been used to study both the interstellar medium of its host galaxy \citep{bowen} and as diagnostics of the wind density \citep{chugai}. For the recurrent nova T Pyx, analysis of the discrete components of this doublet suggested an evolution caused by the progression of a recombination front through ejecta as the optical depth of the ejecta decreases over time \citep{shoreTPyx}. Measurements of the width of the Na\,\textsc{i} D line in supernova impostor SN 2011A were used to determine the origin of the absorption as arising from the CSM \citep{dejaeger11a}.

Besides being used to study the properties of individual objects, the Na\,\textsc{i} D line has been used in a statistical sense to show a correlation that exists between a source's Milky Way extinction and the equivalent width of this doublet \citep{poznanskiextinction}. This is often used as a method of estimating a source's extinction from its spectrum. The doublet has also been used, in the case of SNe Ia, to probe host galaxy reddening. An analysis comparing the equivalent width of the Na\,\textsc{i} D line to the colour excess $E(B - V)$ in a sample of SNe Ia has shown that sources cluster around two lines of significantly different slopes \citep{variety}.

Despite the previous work done regarding this absorption line, there has not yet, to our knowledge, been a systematic investigation into the diversity of Na\,\textsc{i} D evolution for a sample of ILRTs.

We present a method by which the equivalent width (EW) of the Na\,\textsc{i} D doublet can be consistently and accurately measured across a time series of spectra for a particular transient. We apply this to a number of transients in the ILRT class in order to study and compare their behaviour. Additionally, we present a toy model used to predict the evolution of the Na\,\textsc{i} D doublet as ejecta from a central supernova passes through a given density profile of CSM.

In Section \ref{sec:methods} we describe our method of fitting spectral lines using MCMC methods, along with the checks we perform to ensure its accuracy. In Section \ref{sec:model} we introduce our toy model used to predict the evolution of the Na\,\textsc{i} D doublet and show some of these predictions. In Section \ref{sec:results} we present our results from analysing the spectra of our sample of ILRTs. Finally, in Section \ref{sec:conclusions} we summarise our conclusions from this study.

\section{Methods} \label{sec:methods}

\subsection{Line fitting}

The majority of the ILRTs we choose to analyse have spectra publicly available from the WISeREP database\footnote{\url{https://wiserep.weizmann.ac.il}} \citep{wiserep}. In addition to these, we analyse spectra of four ILRTs from \cite{caiILRTs}.

We also checked the ESO archive for any high-resolution spectra taken for any of these ILRTs and found a single high signal-to-noise high-resolution UVES spectrum for NGC 300-2008OT1. We further augment our analysis of NGC 300-2008OT1 with a number of lower resolution spectra from Valerin et al. (in preparation).

A number of our spectra for AT 2013la displayed a broad emission feature at the wavelength of the Na doublet. These spectra were each taken using the same instrument (OSIRIS at the Gran Telescopio Canarias). As we had spectra of the same source from other instruments which did not show this emission feature, we suspect that this effect was instrumental, caused by an inaccuracy in subtracting the background sky spectrum leaving detections of this emission line from the atmosphere. Ideally, we would confirm this through the sky spectra, but these were not available in this case. As such, we discard spectra that show this emission feature as it impacts our ability to fit the absorption from the ILRT accurately.

Five low resolution spectra were available for the ILRT SN 2002bu \citep{2011MNRAS.415..773S} on WISeREP. However, we decided not to include this object in our sample due to a combination of low signal-to-noise and inconsistencies between the spectra making the unambiguous identification of the Na doublet difficult.

In total, our sample consists of ten ILRTs with at least two spectra. We measure the equivalent width of the Na doublet in each of these spectra using a custom-built Markov chain Monte Carlo (MCMC) pipeline using the \textsc{emcee} Python package \citep{emcee}. See \cite{MCMC} for a full review of MCMC methods. A detailed description of the code which we use to perform these calculations is available in Appendix \ref{app:line}.

\subsection{Synthetic spectrum testing} \label{sec:synthhires}

To ensure that this method of fitting the absorption doublet produces reliable results, we test its ability to estimate the equivalent width of the Na doublet from a synthetic line of known strength. We begin by creating a flat, high-resolution continuum with constant flux of 1. We then inject absorption features into this continuum at the expected wavelengths of each component of the doublet. Each component is implemented as a Gaussian with a width of 0.15~{\AA}. This allows the two components to be separated and distinguished individually, which would require a spectral resolution greater than that available from the majority of our real spectra. The amplitude of the D1 component is set to half that of the D2 component.

We then convolve this spectrum with a broad Gaussian whose width is 3~{\AA}, similar to the width of the Na absorption line present in many of our real spectra. This blends the two components of the Na doublet together. Next, we rebin the spectrum such that its resolution matches that of the low resolution spectra. We do this using a method from the PySynphot package \citep{pysynphot} which conserves flux in each bin as we degrade the spectral resolution. Finally we add Gaussian noise proportional to the flux at each wavelength to the spectrum to reduce the signal-to-noise ratio to a level better representative of our real spectra.

At this point, we have a synthetic Na absorption spectrum which resembles the majority of our real spectra with an equivalent width known from our original high-quality spectrum. We generate 1500 synthetic spectra in this described manner, each spectrum differing by the Gaussian noise injected into them. We then use these synthetic spectra to test the effectiveness of our MCMC code by fitting Na doublets to each of them, and comparing these 1500 calculated equivalent widths to the known value from the initial synthetic spectrum.

Each run of the code returns an equivalent width and an associated uncertainty. We record each case where the calculated equivalent width differs from the true value by less than its uncertainty as a successful determination of the line's true strength. We find that the code recovers the expected equivalent width in 68.5 per cent of cases. This value is close to the proportion of a normal distribution which lies within 1 standard deviation of the mean, giving us confidence that our method is recovering equivalent widths at a satisfactory level.

When examining the equivalent widths that our code predicts, we find that our code does have a slight tendency to overpredict rather than underpredict. Overall, our methods calculated a value for equivalent width lower (higher) than the true value 43.3 (56.7) per cent of the time. However, as we can still calculate the true equivalent width within our uncertainty the correct proportion of the time, we do not consider this an important issue. In particular, as we want to track the evolution of the strength of the Na doublet across multiple spectra, a slight systematic bias towards overprediction may manifest in individual spectra but should not negatively effect the overall evolution.

\subsection{IRAF comparison}

\begin{figure}
    \centering
    \includegraphics[width = \linewidth]{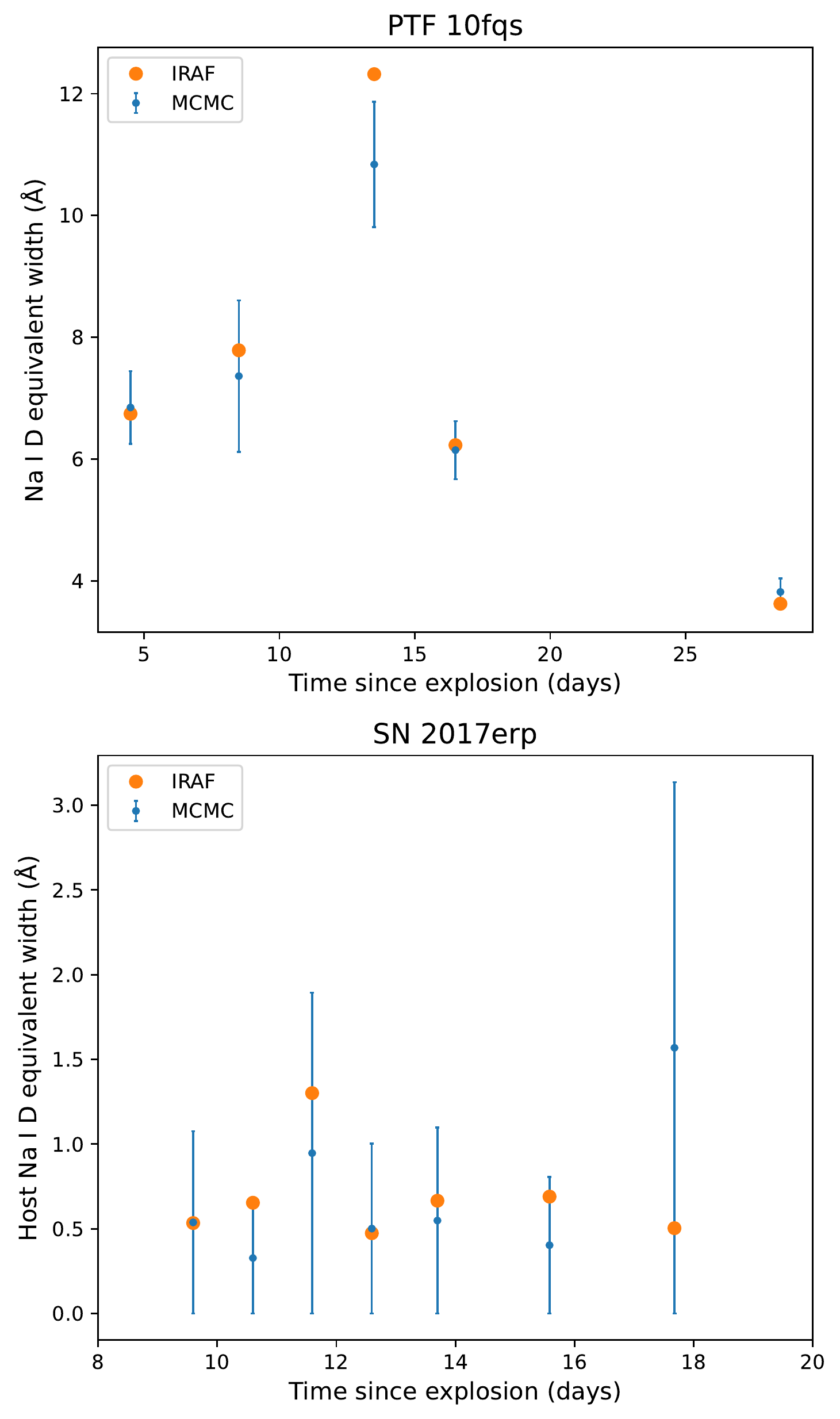}
    \caption{Comparison between the measurements of the Na doublet in PTF 10fqs and the host galaxy of SN 2017erp using our MCMC code and {\sc IRAF}.}
    \label{fig:irafcomp}
\end{figure}

To test the accuracy of our MCMC code, we compare its predictions to those made using the \textsc{IRAF} code \citep{iraf}. We predict equivalent widths using IRAF for a number of spectra for one ILRT: PTF 10fqs, as well as a type Ia supernova which we discuss further in section \ref{sec:17erp}: SN 2017erp. These fits are plotted alongside our calculations using MCMC methods in Figure \ref{fig:irafcomp}.

Within our calculated uncertainties, our results are largely consistent with those from \textsc{IRAF}, with only a single point for PTF 10fqs lying outside of our uncertainties. Our error bars indicate 1$\sigma$ uncertainties, and as measurements from \textsc{IRAF} do not include uncertainties, this discrepancy is small enough to be unimportant. Our measurements of Na in the host galaxy of SN 2017erp all line up with \textsc{IRAF} measurements, although our uncertainties are much larger in this case due to the line being intrinsically weaker as well as this source having a more complicated continuum than most ILRTs.

These comparisons show that our code serves as a method of predicting the strength of line with a similar accuracy to \textsc{IRAF}, with the added benefit of generating uncertainties for each measurement of equivalent width.

\subsection{SN 2017erp} \label{sec:17erp}

In addition to our sample of ILRTs, we examine spectra for a control case. For this, we choose the type Ia supernova SN 2017erp. SN 2017erp is a normal type Ia supernova with some reddening in the near-ultraviolet \citep{2017erp}. We choose this object for a number of reasons. Firstly, it has a number of spectra available on WISeREP, and these spectra show measurable Na absorption lines at redshifts of the host galaxy, as well as from the Milky Way (MW). The object is also at a redshift where host and MW absorption can be distinguished, but is not too far away. Additionally, it is expected that variable Na\,\textsc{i} D lines are uncommon in type Ia supernovae. In a sample of 31 objects where variability could possibly be detected only a single instance of a variable EW was found \citep{Ianovar}. For these reasons, we consider SN 2017erp a good candidate for testing our code, as we expect that absorption from both the host and the MW should remain constant in time.

\begin{figure}
    \centering
    \includegraphics[width = \linewidth]{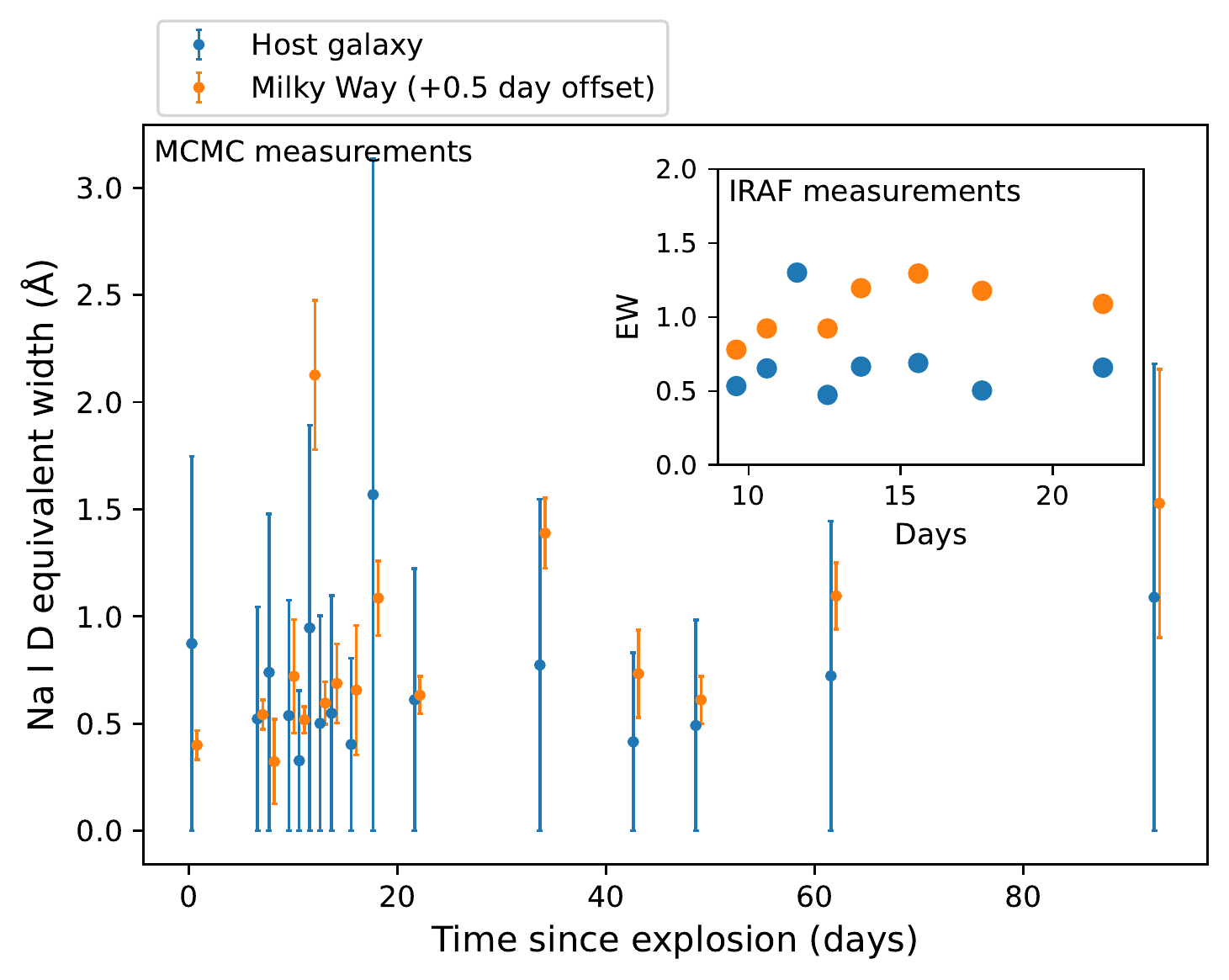}
    \caption{
    Evolution of strength of Na doublet in spectra of SN 2017erp at redshifts corresponding to supernova host galaxy and Milky Way. Main figure shows EW measurements from our MCMC code. Milky Way measurements in main figure are offset by 0.5 days for visual clarity. Inset shows a number of EW measurements produced using \textsc{IRAF}, where correlation between Milky Way and host galaxy measurements is apparent.
    }
    \label{fig:17erpinset}
\end{figure}

We track the strength of both Na doublets across the spectra for SN 2017erp using our MCMC methods. These measurements are shown in Figure \ref{fig:17erpinset}.

As interstellar material is not associated with a variable star or transient, it is expected that its column density should remain static over the time-scales of the observations. Therefore, we expect that the equivalent width of the absorption produced from interstellar material along the line of sight to the source should remain constant in time, regardless of location in either the Milky Way or host galaxy.

In SN 2017erp's host galaxy, our calculated uncertainties are large enough that each measurement is consistent with a non-evolving Na doublet. Measurements from the Milky Way show a larger discrepancy. 

The best agreement is given by an equivalent width of $\sim$ 0.55~{\AA}, for which 9 of 16 measurements (56 per cent) agree within 1$\sigma$. This is slightly lower than the 68 per cent of measurements expected to lie within 1$\sigma$ for normally distributed measurements of a constant value.

We note that the evolution displayed by the Milky Way doublet seems to correlate with the evolution from the host galaxy doublet. In order to examine this potential correlation in closer detail, we fit a number of these spectra between +9 and +23 days using \textsc{IRAF}. These measurements are displayed in the inset of Figure \ref{fig:17erpinset}.

We can see from these measurements that a correlation is present between the strength of the Na doublet in the Milky Way and the strength in the host galaxy. No physical process can explain this correlation, and thus we propose that this supposed variability is not physical in origin, but arises from the spectra themselves. 
In SN 2017erp, the region surrounding the two instances of the Na doublet is dominated by a number of other broad emission and absorption lines, making the fitting of a sensible continuum much more difficult. We propose that the inability to fit a confident continuum to this object affects the measured equivalent widths, and causes this variability. This can result in catastrophic outliers in the calculation of the equivalent width, as seen in the spectrum from +11 days. This also manifests as much larger uncertainties in the equivalent widths.

Overall, we do not expect this effect to have a large impact on our measurements for ILRTs. The variability seen in the measurements from SN 2017erp is far smaller than those we measure in our sample of ILRTs. Additionally, ILRT continua tend to be much flatter and less complex around the Na absorption line, with very few additional lines impacting on our measurements. This allows us to calculate a continuum and corresponding equivalent width with more certainty.

Ideally, we would measure the equivalent width of Na at zero redshift in each ILRT spectrum. This would allow us to make measurements on a quantity which should remain constant using spectra with simple continua. This could be used as a check that the evolution measured at host redshift is physical in origin. Unfortunately for all the ILRTs in our sample the redshift is either small enough that the Milky Way absorption is blended with that of the host, leaving us either unable to distinguish the two or, where there is separation between the two lines, Milky Way absorption is too faint to be measured.

\section{Toy model} \label{sec:model}

In order to guide our expectations for the evolution of the Na\,\textsc{i} doublet, we create a simple toy model. A number of models have been developed to explain the variability of Na absorption, particularly in SNe Ia. \cite{patat06x} suggest a model for SN 2006X where variability arises from the ionisation and subsequent recombination of Na in the CSM. \cite{borkowski} extend this by considering the level of photoionisation as a function of the location of CSM shells around the central supernova. \cite{soker} offers a model where the Na responsible for absorption is released from dust grains through photon-stimulated desorption.

Our toy model is based on an ejecta front from a central supernova travelling outwards at constant velocity through a spherically symmetric region of circumstellar material. We track the amount of Na\,\textsc{i} present in the CSM in front of the ejecta and use this as a proxy for the strength of the Na\,\textsc{i} D absorption feature.

Using column density of CSM as a proxy for the equivalent width of the Na\,\textsc{i} D line relies on the assumption that no saturation of this line occurs. In such a case, the two quantities may not correlate with one another simply. In our analysis of real ILRT spectra, we found that the absorption doublet was never strong enough to come close to saturation, and so we proceed with this assumption.

This model is similar conceptually to the \cite{borkowski} model, although in a different regime in terms of scale and density. Their model focuses on bubbles on parsec scales blown into the interstellar medium by strong accretion-driven winds from the progenitor of a Type Ia supernova, and hence the CSM will not be overrun by ejecta on the timescales of interest to us. In contrast, our model relates to the denser circumstellar material much closer to the star. 

We first select a density profile for the circumstellar material. The main density profiles of interest are that of a typical stellar wind proportional to $r^{-2}$, and a constant density profile, though any profile can be used.

We assume that at time t = 0 directly after the explosion, all Na within the CSM has been ionised to Na\,\textsc{ii} by emission from the central supernova. Na has a relatively low ionisation potential of 5.139 eV, corresponding to a photon wavelength of 241 nm, so the strong radiation field of the supernova should be sufficient to produce such a scenario.

In addition to ionising the Na, it is likely that the supernova will ionise a significant fraction of hydrogen in the CSM. As the relative fraction of H to Na in the CSM is high, this will result in a large amount of free electrons being present. These free electrons will allow for the recombination of Na\,\textsc{ii} to Na\,\textsc{i}. Due to the high density of free electrons compared to that of Na, we assume that Na\,\textsc{ii} recombines at a rate proportional to its density, i.e. at each time step some fixed percentage of the remaining Na\,\textsc{ii} recombines to Na\,\textsc{i}.

Starting at r = 0, we move the position of the ejecta front outwards through the CSM at a constant velocity with each timestep. At each step, we then calculate the total column density of Na remaining in front of the shock by integrating the density function past that point. We then let a fixed proportion of the Na\,\textsc{ii} recombine to Na\,\textsc{i}, and multiply this fraction by the total column density in order to estimate the column density of Na\,\textsc{i} in front of the ejecta. We repeat this process and track the evolution of the column density of Na\,\textsc{i} as a proxy for the strength of the Na\,\textsc{i} D doublet.

\begin{figure*}
    \centering
    \includegraphics[width = \linewidth]{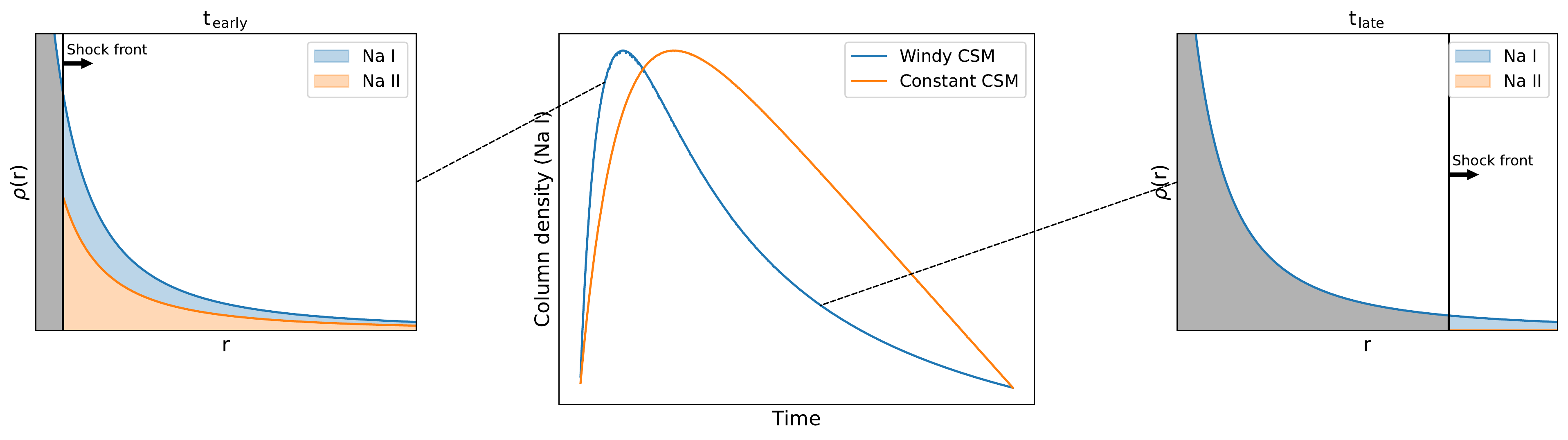}
    \caption{Evolution of column density of Na\,\textsc{i} in front of constant velocity ejecta passing through circumstellar material with wind-like ($\propto r^{-2}$) and constant density profiles. This quantity is used as a proxy for the strength of the Na\,\textsc{i} D absorption doublet. Axes are given in arbitrary units and wind-like and constant CSM evolutions are normalised to the same peak.}
    \label{fig:evolution}
\end{figure*}

In Figure \ref{fig:evolution}, we display the predictions from our toy model for the evolution of the column density of Na\,\textsc{i} given ejecta sweeping through two circumstellar environments. The first has a density structure proportional to $r^{-2}$, typical of a stellar wind, while the second displays a constant density CSM. As we expect the strength of the Na absorption doublet to track this quantity, these can be interpreted as predictions for the evolution of the equivalent width of the Na doublet over time.

The central panel of Figure \ref{fig:evolution} shows the integrated column density of Na\,\textsc{i} remaining in front of the shock as time progresses for either density profile. The panels on either side display the structure of the CSM in the windy regime at two representative times, one early and one late. In these panels, the black vertical line represents the position of the shock travelling outwards through the CSM, with the grey shaded region representing the CSM left behind it. Blue and orange shaded regions in front of the shock represent the relative proportions of Na\,\textsc{i} and Na\,\textsc{ii} in the unshocked CSM.

At early times, it can be seen that a large amount of CSM remains in front of the shock, split relatively evenly between both states of Na, resulting in the production of a strong absorption feature. At late times, although most of the Na\,\textsc{ii} has recombined to form Na\,\textsc{i}, there is comparatively little left in front of the shock, resulting in a weak absorption feature.

As we are largely concerned with the shape of evolution, we display these plots without units, and normalise the curves to the same peak. It can be seen that both profiles of CSM produce a similar overall shape in their evolution, where the column density rises to a peak before dropping down to a low level. The windy CSM shows a faster rise followed by a concave decrease over time, levelling out at a low strength. The constant density CSM rises slightly slower, and decreases from peak at a more linear rate.

\section{Results} \label{sec:results}

\subsection{ILRT evolution}

\begin{figure*}
    \centering
    \subfloat[]{
      \includegraphics[width=0.32\linewidth]{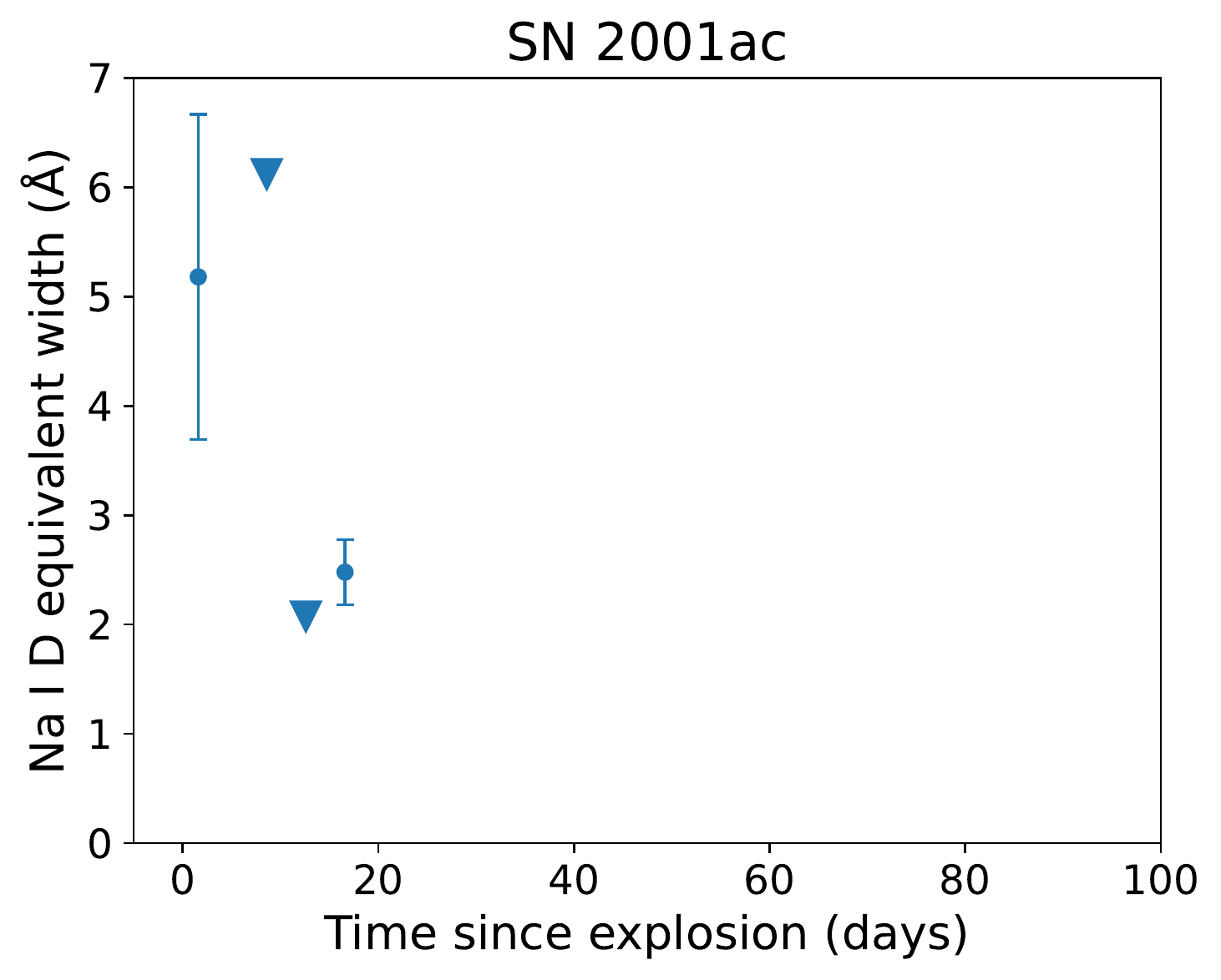}
    }
    \subfloat[]{
      \includegraphics[width=0.32\linewidth]{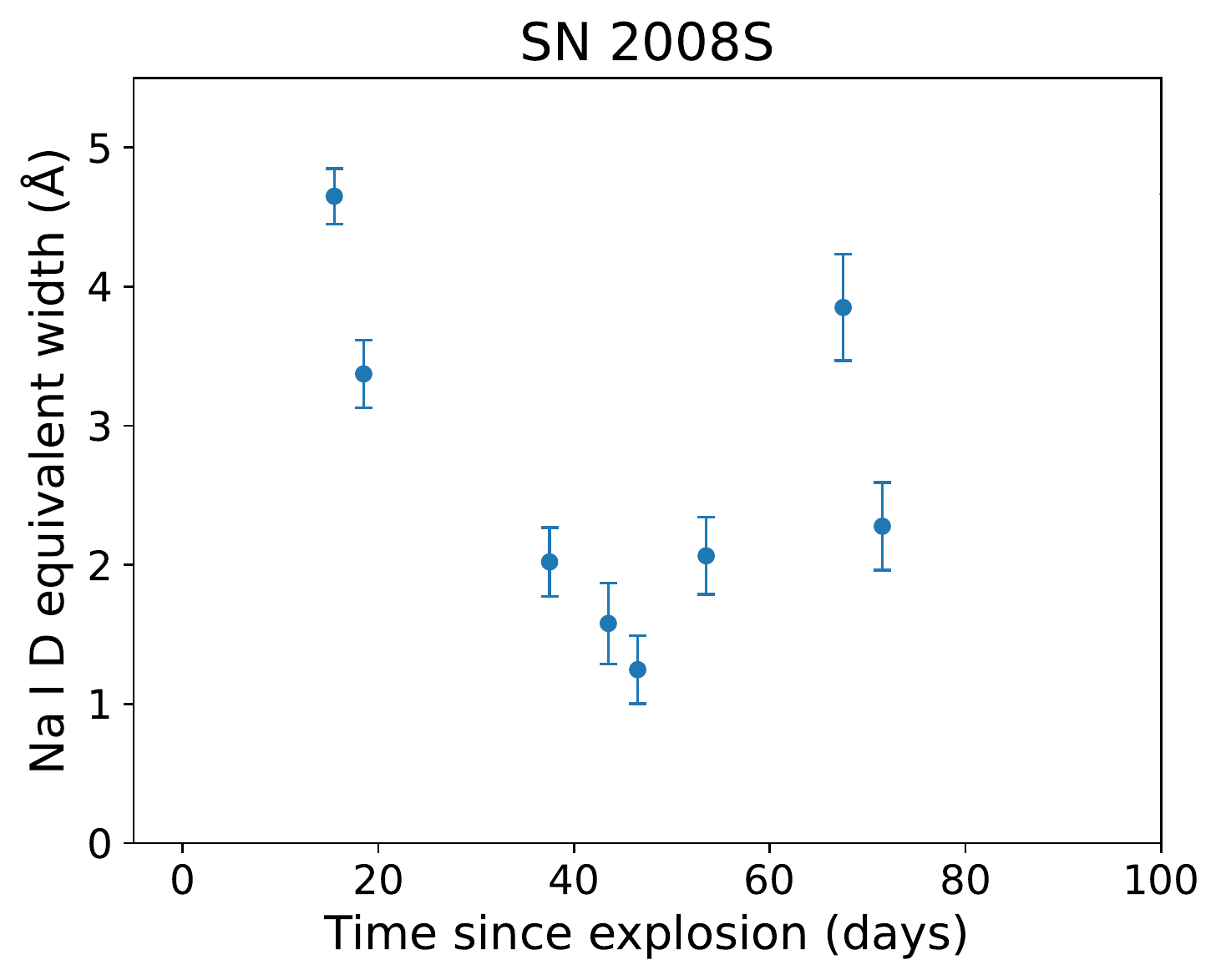}
    }
    \subfloat[]{
      \includegraphics[width=0.32\linewidth]{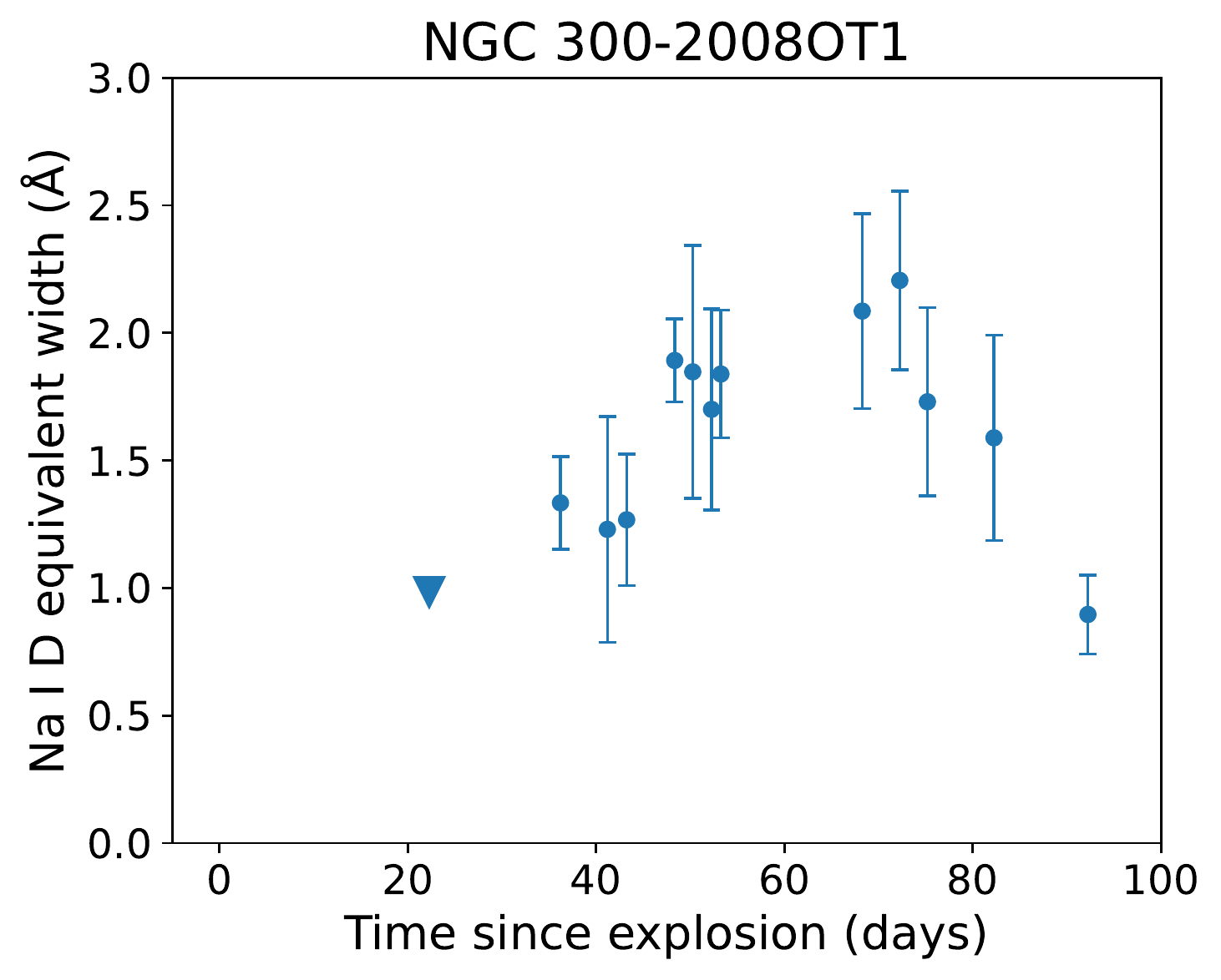}
    }
    
    \hspace{0mm}
    
    \subfloat[]{
      \includegraphics[width=0.32\linewidth]{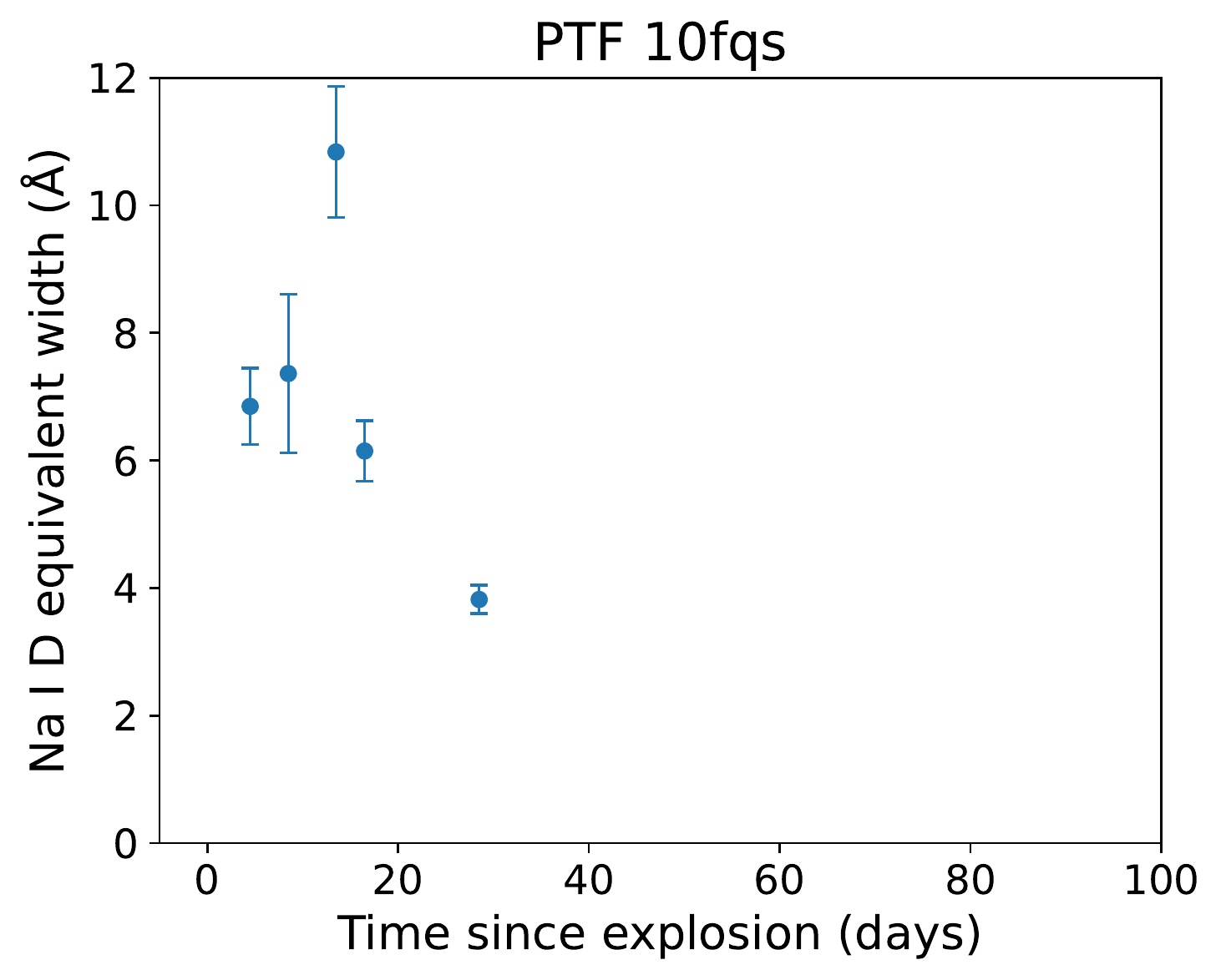}
    }
    \subfloat[]{
      \includegraphics[width=0.32\linewidth]{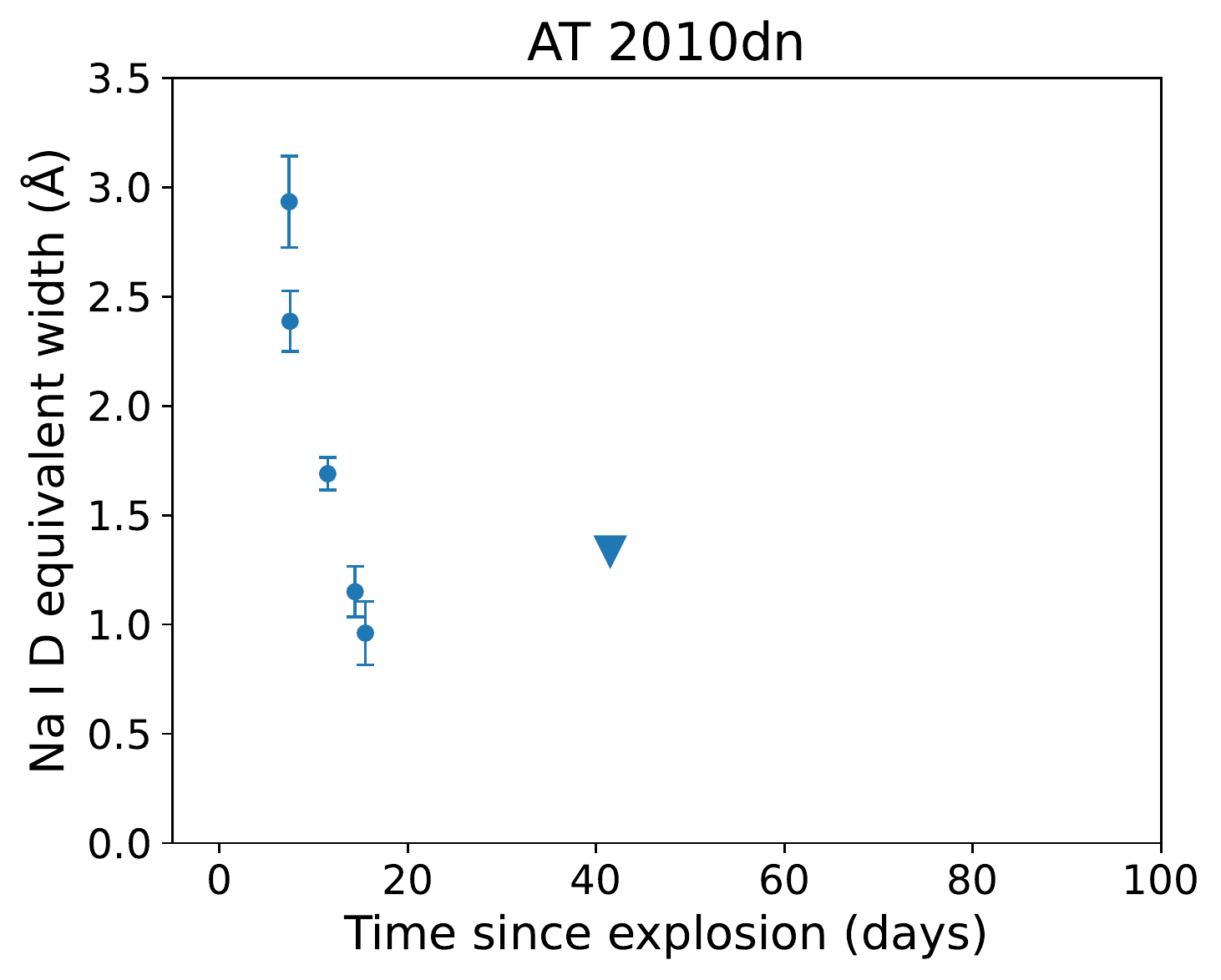}
    }
    \subfloat[]{
      \includegraphics[width=0.32\linewidth]{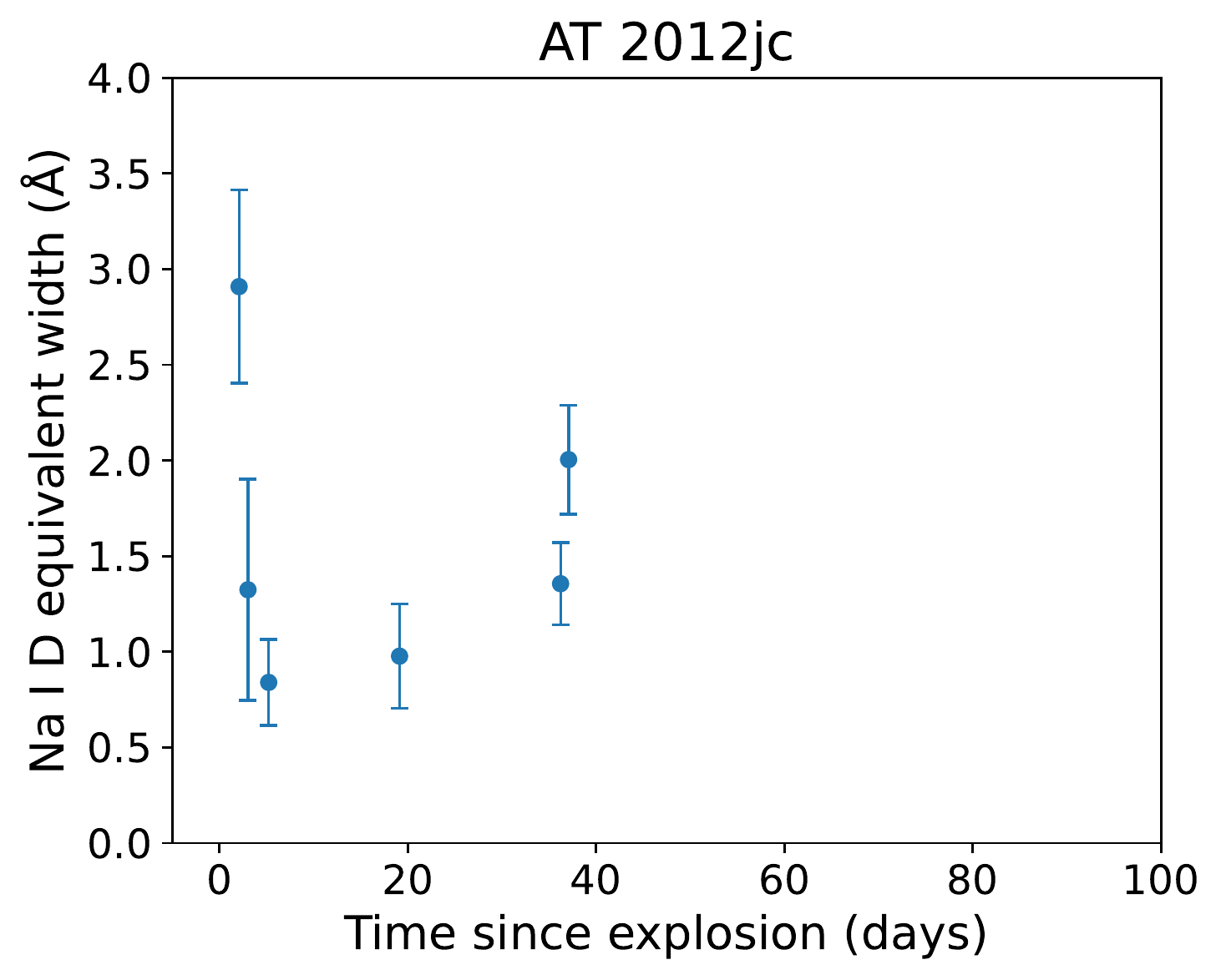}
    }
    
    \hspace{0mm}
    
    \subfloat[]{
      \includegraphics[width=0.32\linewidth]{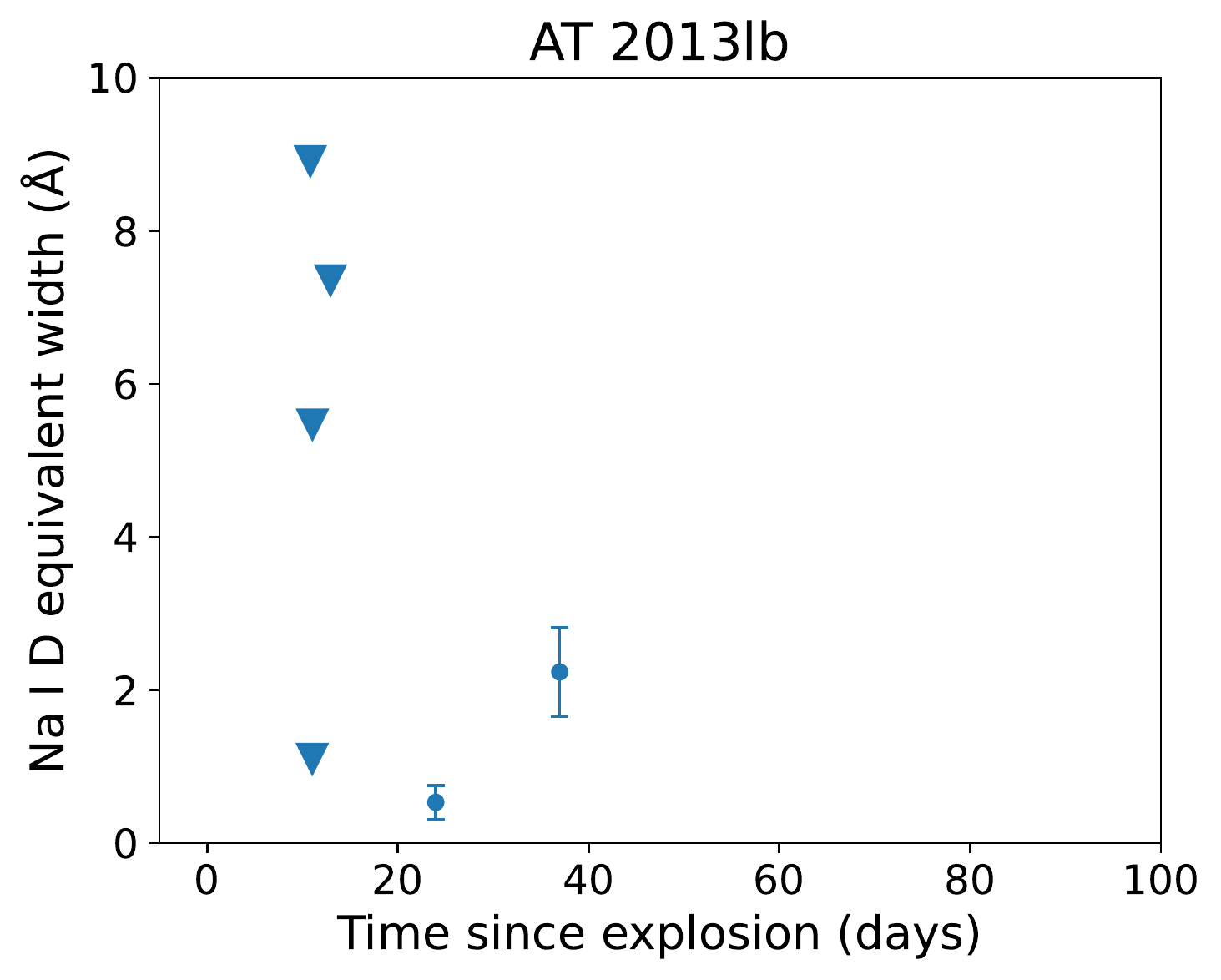}
    }
    \subfloat[]{
      \includegraphics[width=0.32\linewidth]{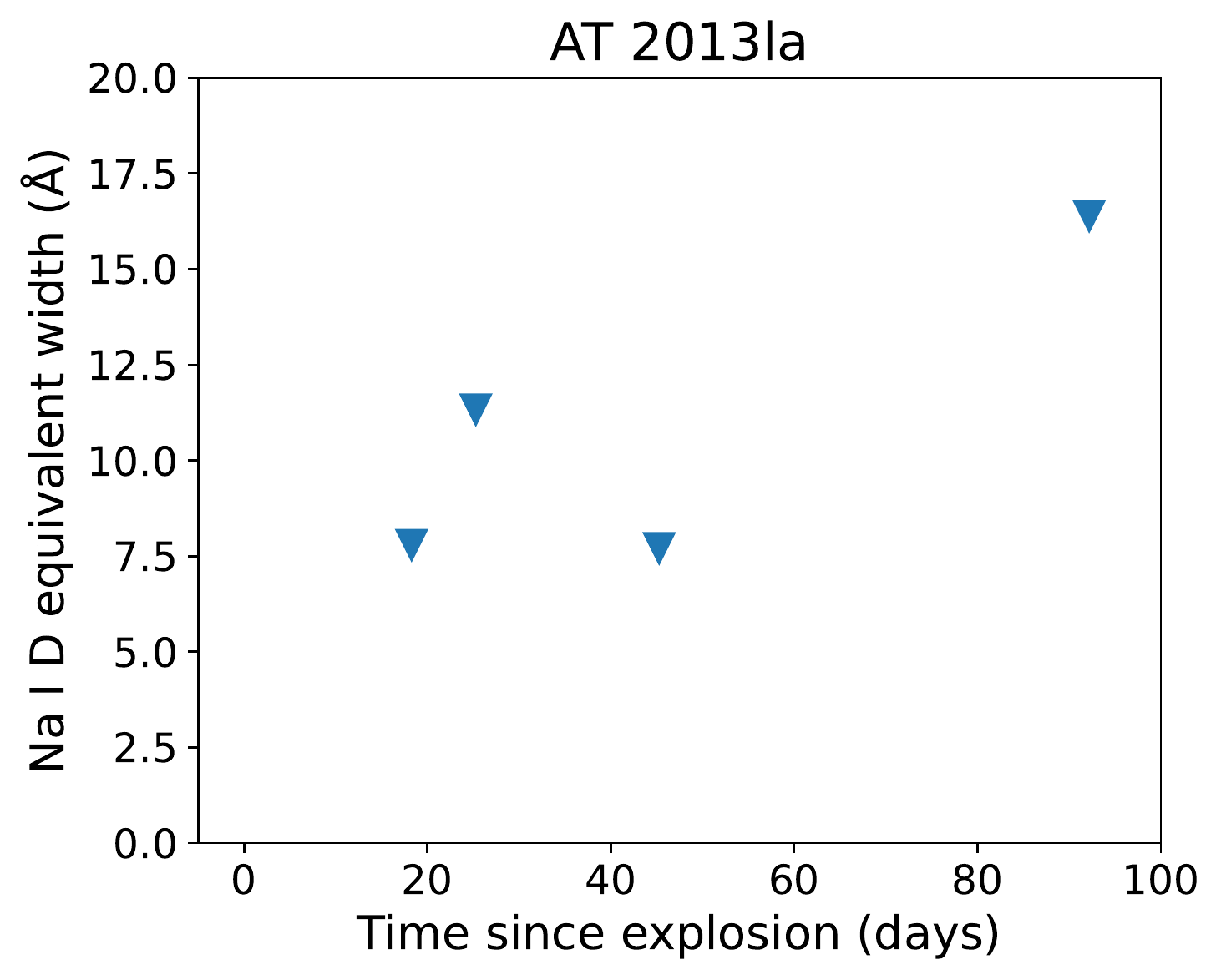}
    }
    \subfloat[]{
      \includegraphics[width=0.32\linewidth]{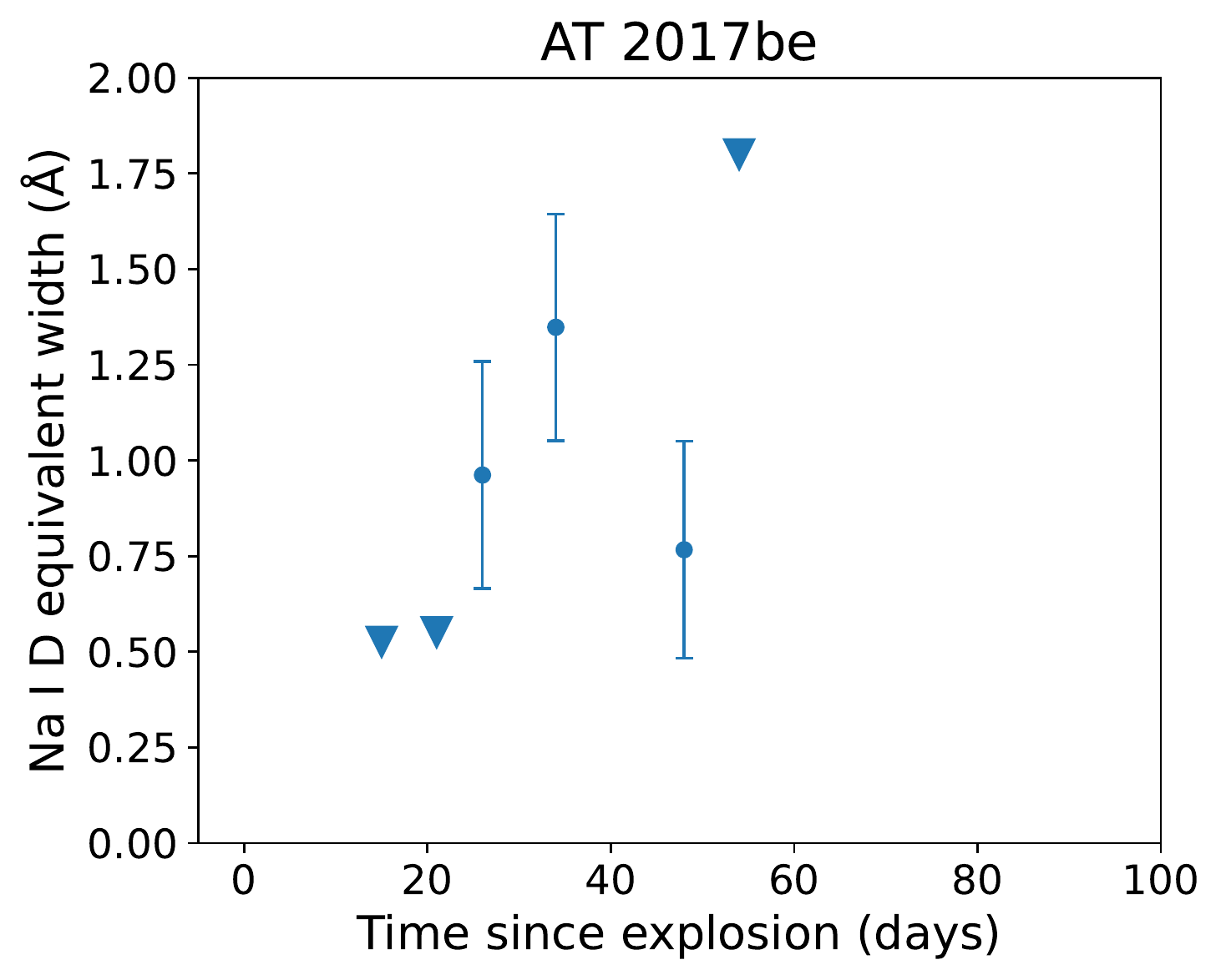}
    }
    
    \hspace{0mm}
    
    \subfloat[]{
        \includegraphics[width=0.32\linewidth]{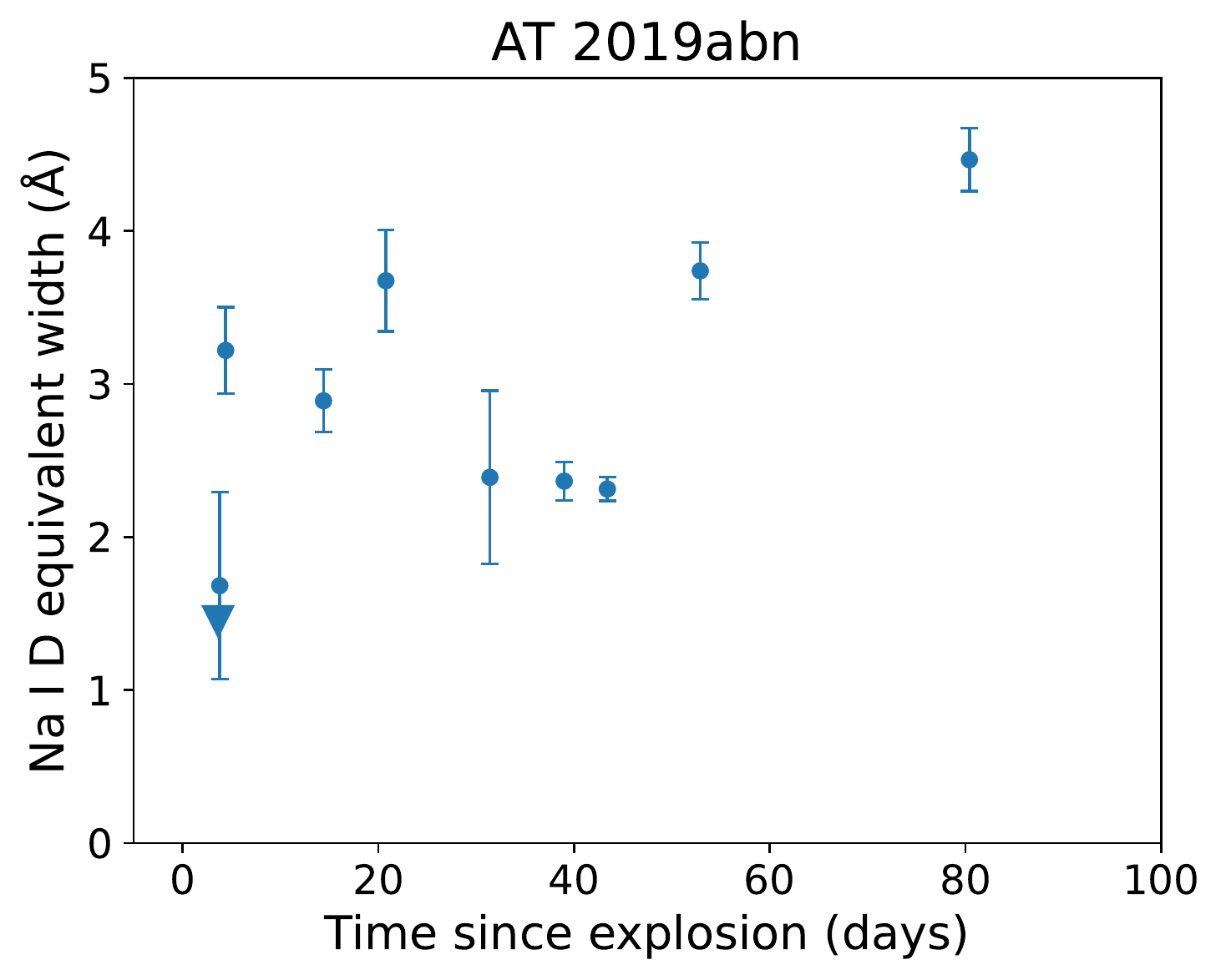}
    }

    \caption{Evolution of the equivalent width of the Na\,\textsc{i} D absorption doublet in our sample of ILRTs. Each plot displays the evolution for the first 100 days post-explosion, though the scales for equivalent width differ for each plot. Individual descriptions for the evolution of each ILRT are given in the main body of text. Triangles represent upper limits to equivalent width in non-detections.}
    \label{fig:realEWs}
\end{figure*}

Figure \ref{fig:realEWs} shows the evolution of the strength of the Na\,\textsc{i} D absorption line as measured using our MCMC code for our sample of ILRTs.

Below we provide brief descriptions of the evolution displayed by each of the ILRTs. Phases are given in terms of days from explosion where one is available in the literature, or days since first detection where no explosion epoch is available.

\begin{enumerate}[label=\textbf{\alph*}), labelwidth = 0pt]
\item \textbf{SN 2001ac}

Only four spectra are available for SN 2001ac \citep{2011MNRAS.415..773S, 2017PASP..129e4201S}, and Na is only detected in two of these. Our first detection is at +1.6 days, which shows a strong absorption line but with large uncertainties. This is followed by two non-detections, the first of which constrains the equivalent width to the same strength as the previous detection or lower. The second non-detection is around the level of our second detection at +16.6 days, which shows the absorption becoming weaker with time.

\item \textbf{SN 2008S}

Detections of the Na doublet in SN 2008S \citep{botticella08s} show evidence of long-term evolution. Our first detections show the strength of the doublet decreasing over time, from +18.5 days until +46.5 days, where it begins to increase again until a peak at +67.5 days. One further detection at +71.5 days shows the line decreasing in strength rapidly. One non-detection is not shown at +101.5 days, however the upper limit for this non-detection is not strongly constraining, at 5.32 {\AA}.

\item \textbf{NGC 300-2008OT1}

We have no early observations for NGC 300-2008OT1 (Valerin et al., in prep.), with our first initial limiting constraint on the strength of the absorption at +22.3 days. After this the absorption feature appears and grows in strength until roughly 70 days, before declining again. The amplitude of this variation is smaller than that seen in many of the ILRTs in our sample. Interestingly, the shape of this evolution showing a rise to single peak followed by a decline is similar to the behaviour predicted by our toy model. In our models, the strength of the Na line begins to increase immediately post-explosion and begins to decrease when the shock has passed through a large enough amount of CSM such that the decrease in overall column density outmatches the increase in the fraction of Na\,\textsc{i} in the CSM. Observations sooner after explosion would allow us to determine whether the Na line is increasing in strength uniformly from the time of explosion, however these observations would require a high signal-to-noise ratio in order to detect such a weak line ($<$~1~{\AA} at +22.3 days).

The late time-scale of this evolution may indicate that it is being driven by interaction with CSM far from the central supernova. We fit a simple quadratic curve to our detections and find that this curve peaks and begins to decline at $\sim$~+66 days. We make a rough estimate for the ejecta velocity from an ILRT. Ejecta from typical core-collapse supernovae tends to travel at velocities on the order of $10^4$~km~s$^{-1}$. As ILRTs are expected to be produced by weaker explosions, we expect a lower ejecta velocity than this. In the case of the underluminous SN IIP SN 2005cs, \cite{2005cs} measure an ejecta velocity of $\sim$~1000~km~s$^{-1}$. We consider this velocity to be similar to that of an ILRT, and as such we take our ejecta velocity to be 2000~km~s$^{-1}$ as a conservative estimate for a weak explosion. Assuming that this velocity is maintained for 66 days, this would place the CSM region $\sim 1.14 \times 10^{15}$ cm from the central supernova.

\item \textbf{PTF 10fqs}

PTF 10fqs \citep{kasliwalLRNe} shows a fast rise to a strong peak in absorption, followed by a similarly fast decline in strength with no observations at late times. This evolution looks quite similar to the behaviour expected by interaction with CSM in a windy regime. Unlike NGC 300-2008OT1 above, it appears to begin its rise in strength directly after explosion, and additionally reaches a much larger equivalent width. This may indicate interaction with a denser CSM much closer to the supernova.

\item \textbf{AT 2010dn}

AT 2010dn \citep{caiILRTs} shows a uniformly declining strength of its Na absorption line over time between +7.4 days and +14.4 days, with a single non-detection coming later at +42.4 days. As we have no detections prior to +7.4 days it is possible that there is a rise to this peak beforehand which would look similar to our models, though this would require CSM very close to the supernova.

\item \textbf{AT 2012jc}

AT 2012jc \citep{2012jc, caiILRTs} begins with behaviour similar to that of AT 2010dn, however beginning at a much earlier time. Our first detection comes at +2.1 days. After a very sharp decline in strength, the Na line appears to rebrighten. It should be noted that our first detection comes with a relatively large uncertainty, and in the absence of this point it is possible that no such drop in equivalent width would be inferred. Our toy model assumes that the Na equivalent width should be proportional only to the amount Na\,\textsc{i} remaining in front of the ejecta, with the entirety being ionised to Na\,\textsc{ii} at t = 0. For this model, it is not possible to explain a rebrightening after the strength of absorption begins to decline, as this would require a process by which new Na\,\textsc{i} is somehow injected into the system in front of the ejecta.

One possibility is that the first detection is inaccurate, and there is no drop in equivalent width at early times. This would allow for the remainder of the evolution to be interpreted in line with our toy model. However, as a potential rebrightening behaviour can also be seen in SN 2008S, we suggest a mechanism by which this can occur.

We suggest that these cases may be explained by a different CSM density profile. Instead of being wind-like, we propose that the CSM has a low density closest to the star, before increasing to a maximum some distance from the star and then falling off similarly to that of a wind-like CSM profile. In addition, we suggest that the sodium present in the CSM is not fully ionised by radiation from the initial supernova. This leads to a situation where, at $t = 0$, the majority of Na is un-ionised Na\,\textsc{i}. As the ejecta passes through this nearby low-density region of CSM, the column density of Na\,\textsc{i} decreases, causing the equivalent width to decrease in strength. Eventually the ejecta reaches a region of high density CSM some distance from the star. At this point, the ejecta interacts strongly with the CSM, producing X-ray photons which help to photoionise the outer region of CSM, causing a further drop in Na\,\textsc{i} equivalent width. As the ejecta moves past this region of high density CSM, the strength of interaction decreases, producing fewer ionising photons and allowing for recombination to increase the relative level of Na\,\textsc{i} in the remaining CSM. If the rate of recombination is high enough, this process increases the column density of Na\,\textsc{i} enough to produce a peak in the equivalent width. As the ejecta moves further out into the low-density material, recombination continues until most of the Na is un-ionised, however at this point the ejecta has passed enough that the total column density continues to decrease, as seen particularly clearly in SN 2008S.

\item \textbf{AT 2013lb}

Observations of AT 2013lb \citep{caiILRTs} begin with four non-detections of Na spanning a short period of time from +10.8 to + 12.9 days. Three of these non-detections do not provide strong constraints on equivalent width, but the spectrum at +11 days constrains the strength of the doublet to below 0.72 {\AA}. After this, we see two detections where the strength of the Na line increases over time. Without further observations, we are unable to determine at what time the strength eventually peaks, or if any other behaviour is displayed.

\item \textbf{AT 2013la}

AT 2013la \citep{caiILRTs} is unique among our sample of ILRTs in that we do not find any spectra where the Na absorption feature is detected. The observations we do have are sparsely populated, with no observations before +18.3 days and no observations between +45.3 and +92.2 days. Additionally, the limits on the equivalent width obtained from these non-detections are not very constraining, with many detections of our sources having detections at equivalent widths weaker than any of the limits for AT 2013la. This means that it is possible that Na was present in this source, but was simply undetected, or that the Na doublet displayed fast early evolution, similar to that of AT 2010dn, or late time evolution similar to NGC 300-OT20081. Either of these may have been missed due to the paucity of observations. However, from these observations we can only conclude that there is currently no evidence for the Na\,\textsc{i} doublet appearing in AT 2013la.

\item \textbf{AT 2017be}

Observations of AT 2017be \citep{cai17be} begin with two non-detections which constrain the strength of the Na doublet to a very weak level, below $\sim$~0.5 {\AA}. Soon after these we see three weak detections which are consistent with one another within uncertainties. These uncertainties are relatively large compared to the strength of the line. After these detections we have one further non-detection which places a relatively loose constraint on the equivalent width. Without further detections we cannot determine whether the doublet remains at the same level as seen in the initial three detections, or begins to increase or decrease in strength.

\item \textbf{AT 2019abn}

AT 2019abn \citep{jencson19abn} shows a large amount of variation in its Na line strength over time. It begins by increasing in strength, before dropping down to a lower level by $\sim$ +40 days, after which it increases once more, remaining at a strong level by the final observation. As mentioned in our description of AT 2012jc, our models can explain an increase in strength followed by a decline, but are not able to explain a late rebrightening of the Na line.

\end{enumerate}

As can be seen, our sample of ILRTs shows a diverse range of behaviours in terms of strength, time-scale, and shape of evolution. While we do not see the same type of evolution displayed across the entire class, we do see a number of similar patterns in a few of these objects.

Both PTF 10fqs and NGC 300-2008OT1 show an evolution similar to that predicted by our toy model with a single peak, although they differ widely from one another in terms of the strength of the line and time-scale. This evolution is also, qualitatively speaking, the most similar to those predicted by our toy model of ejecta passing through a recombining region of CSM.

A number of ILRTs seem to show the strength of the Na line increasing after some time, some of these after an initial period where it decreased in strength. These include SN 2008S, AT 2012jc, AT 2013lb, and AT 2019abn.

Further ILRTs show their own individual evolution, including the non-detection in AT 2013la. AT 2017be in particular shows only three weak detections with consistent equivalent widths. As previous observations showed non-detections for this ILRT we can say that the Na line is varying in time, but it remains to be seen what type of evolution it follows after these two detections.

One key insight is the diversity of time-scales on display in these evolutions. Observations of these ILRTs are sparse and unevenly sampled, meaning that some of the outlying ILRTs which seem to show their own unique evolution may in actuality appear more similar to others if further observations were available. This highlights the need for further spectroscopic coverage of future events in order to better constrain the evolution of this line.

We find that all but one of our ILRTs display clear evidence of variability in the Na\,\textsc{i} D line over time. In our one exception, AT 2013la, we find no detection of the Na line. This does not necessarily rule out the possibility of Na evolution in this ILRT, it may be that the variability was simply missed due to sparse observations, or the amplitude of the line may have been low. Indeed, our limiting equivalent widths for this ILRT are higher than many of our detections for other sources. However, our observations cannot prove that this evolution exists.

Our prior analysis of SN 2017erp was intended to show the ability of our code to distinguish between sources displaying true variability in the Na line and those not undergoing evolution. Although this attempt was complicated by the systematic variability found in the measurements of SN 2017erp, we note that the variability found for SN 2017erp is smaller than that displayed in our sample of ILRTs. Hence, we believe that the evolution displayed by these ILRTs is physical, rather than systematic in origin.

\subsection{NGC 300-2008OT1}

\begin{figure}
    \centering
    \includegraphics[width = \linewidth]{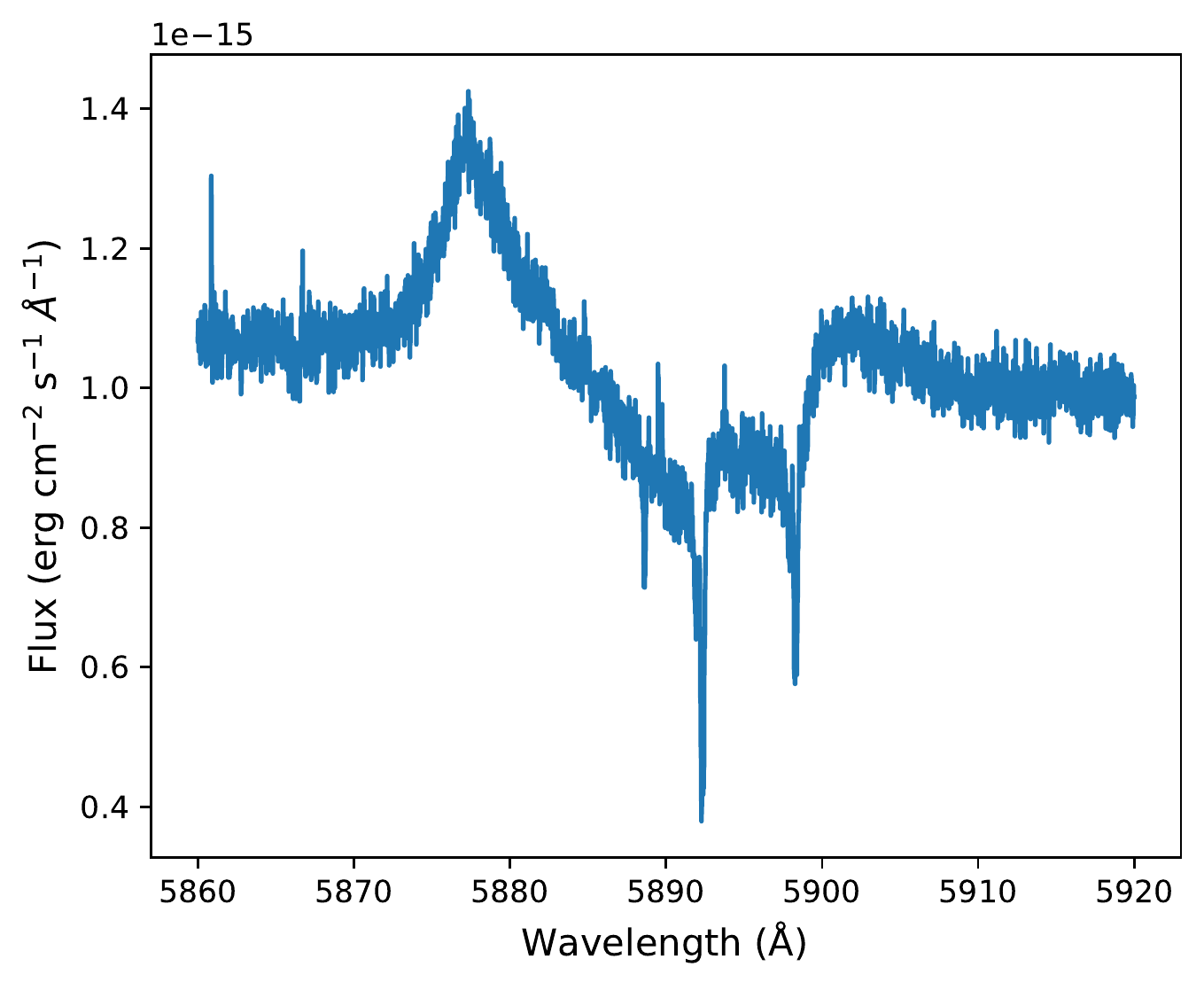}
    \caption{UVES spectrum of NGC 300-2008OT1 at +69.1 days. Broad components of the Na absorption line can be seen in the centre, along with a number of narrow components. He I emission line lies at the blue end of the Na absorption feature.}
    \label{fig:300otoriginal}
\end{figure}

One ILRT in our sample, NGC 300-2008OT1 has an additional high-resolution UVES spectrum. The reduced spectrum is available from the ESO Science Archive\footnote{http://archive.eso.org/cms.html}. In this spectrum, there was clear separation between the D2 and D1 components of the Na doublet, and individual substructure could be analysed for each. The spectrum shows an emission line from He\,\textsc{i} at $\sim$ 5875 {\AA}, followed directly by a broad absorption feature at the wavelengths of the Na doublet. Within this broad absorption feature, further individual narrow absorption features can be seen. We show the original high-resolution spectrum in Figure \ref{fig:300otoriginal}.

\begin{figure}
    \centering
    \includegraphics[width = \linewidth]{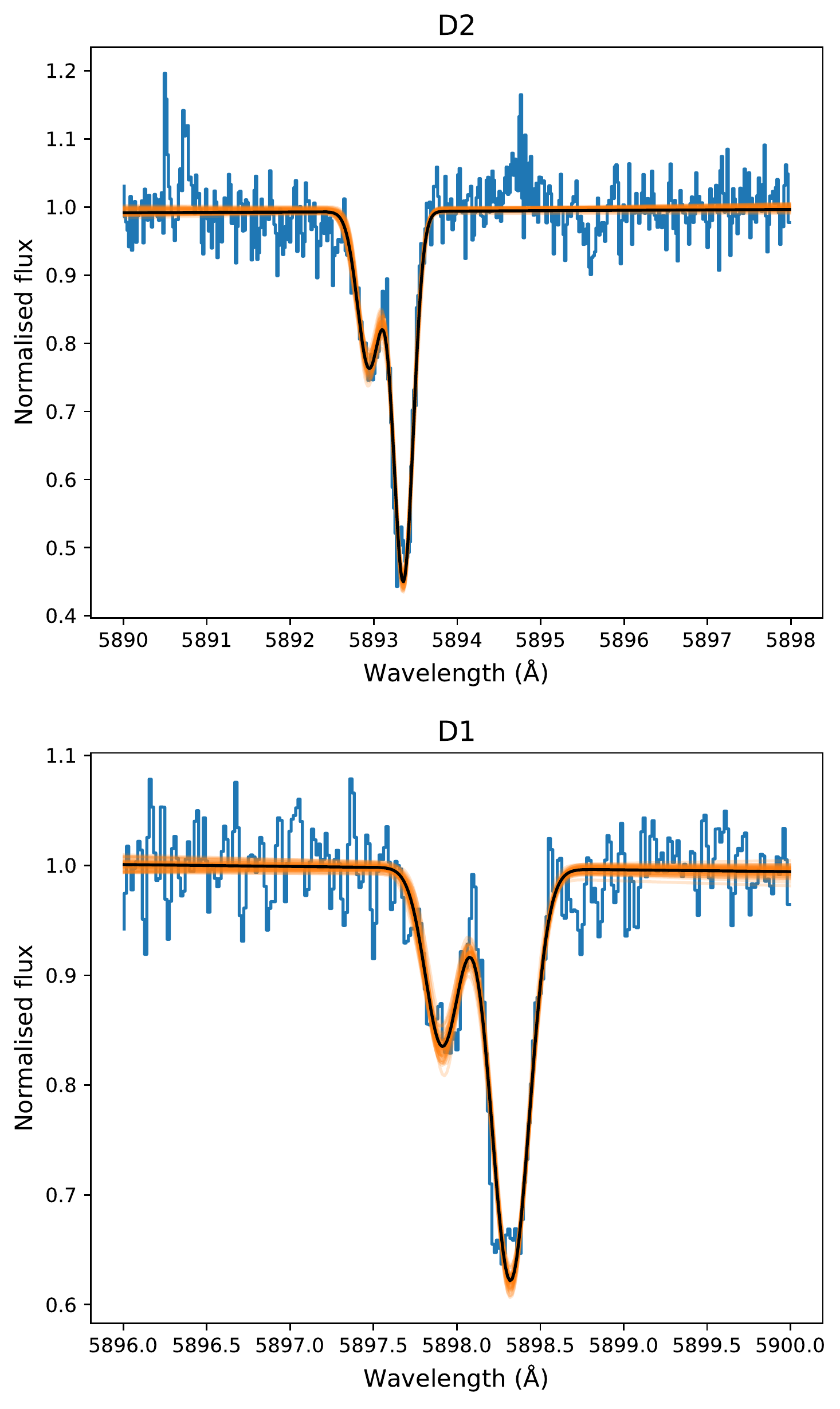}
    \caption{Subsections of the UVES spectrum of NGC 300-2008OT1 centred on the narrow D2 and D1 components of the Na absorption line. In each case, two absorption components can be seen. Black line indicates best fit from MCMC model, while orange represents sample of fits from Markov Chain.}
    \label{fig:zoomed300ot}
\end{figure}

We begin our analysis of this spectrum by fitting the narrow absorption components of Na. Figure \ref{fig:zoomed300ot} shows cutouts of the narrow line structure visible at the wavelengths of the D2 and D1 components. At each position we find two narrow absorption features, which we fit using a slightly modified version of our MCMC models described above, specifying narrower ranges for central wavelengths and widths of each line. We fit the substructures of the D2 and D1 components separately. We attempted fitting three absorption features to these cutouts, and found that this did not improve the accuracy of our fits.

At the D2 position, we measured narrow absorption features at wavelengths of 5892.944~$\pm$~0.011~{\AA} and 5893.354~$\pm$~0.004~{\AA}, with full widths at half maximum corresponding to velocities of 16.44~$\pm$~2.19~km~s$^{-1}$ and 14.14~$\pm$~0.49~km~s$^{-1}$ respectively. At the D1 component, we measured these features at wavelengths of 5897.913~$\pm$~0.008~{\AA} and 5898.321~$\pm$~0.003~{\AA}, with velocities of 12.21~$\pm$~0.99~km~s$^{-1}$ and 13.52~$\pm$~0.40~km~s$^{-1}$ respectively.

Within uncertainties, the velocities of the second components measured at both the D2 and D1 positions are consistent, as would be expected if these features were produced in the same environments. The velocities of the first components differ at a significance of 1.3$\sigma$. This small discrepancy may arise from difficulty in normalising the spectrum as this specific component is located close to the red edge of the broad Na absorption feature.

While these low velocities may be indicative of being produced in two separate regions of slow moving material in the interstellar medium along the line of sight to the source, these velocities are also similar to that of a red supergiant wind. These stars have winds typically on the order of a few 10s of km~s$^{-1}$. We expect similar wind speeds from a SAGB star, and as such we cannot distinguish between these two possible regions as the site of origin of these narrow lines using a single high-resolution spectrum. Time series of high-resolution spectra may be required to confirm either of these options.

\begin{figure}
    \centering
    \includegraphics[width = \linewidth]{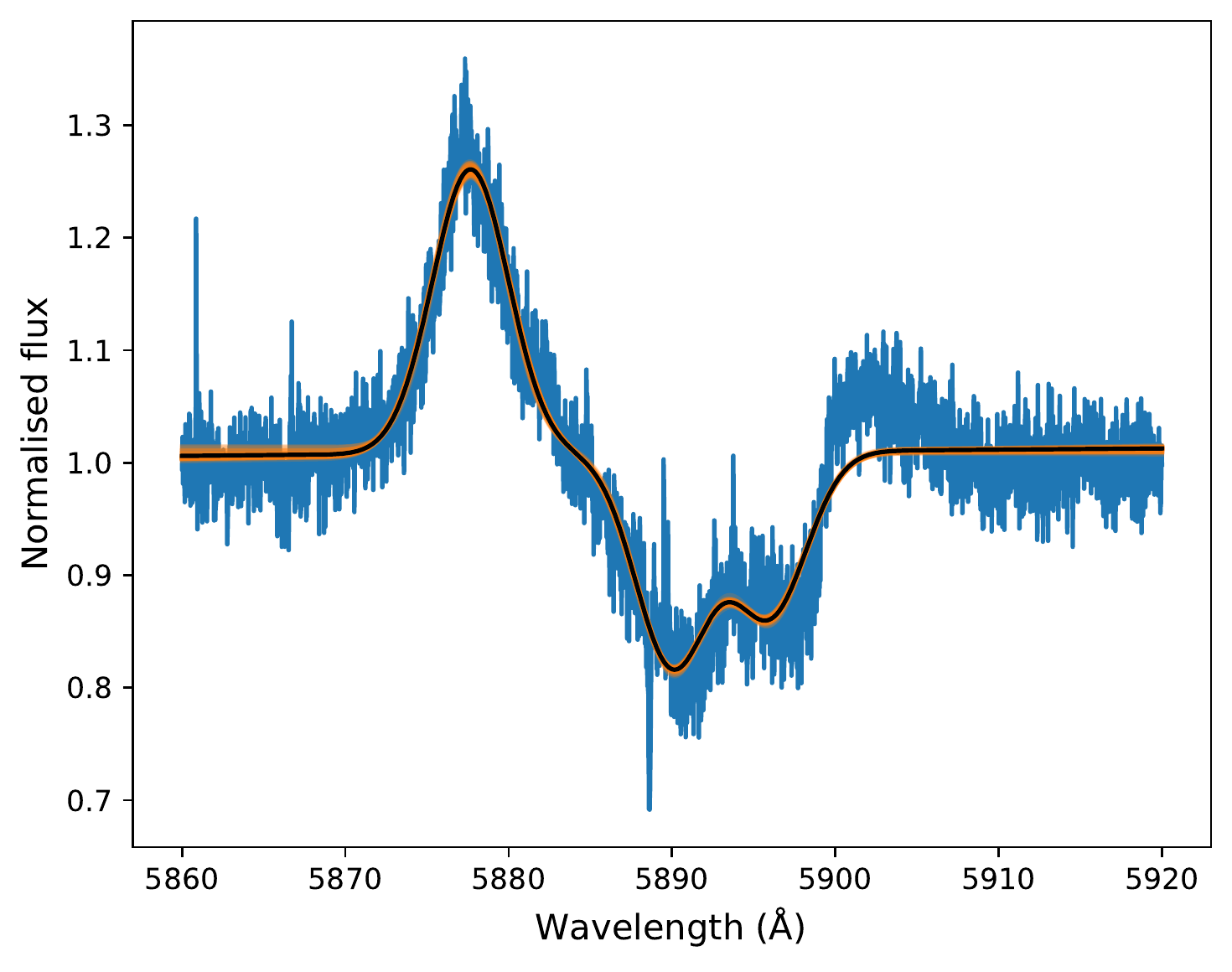}
    \caption{UVES spectrum of NGC 300-2008OT1 at +69.1 days. Narrow components of Na doublet have been removed in order to fit the broad features. Broad absorption from the Na doublet can be seen, as well as the nearby He\,\textsc{i} emission line. Black line indicates best fit from MCMC model, while orange represents sample of fits from Markov Chain.}
    \label{fig:300otspec}
\end{figure}

Next, we fit the broad components of Na absorption visible in the spectrum. We do this by first manually removing the narrow components described above which would interfere with the accuracy of fitting the broad components. To do so, we isolate small regions surrounding these narrow lines in our spectrum and remove them. We then refill these regions using linear fits interpolated between the edges of the removed portions of the spectrum. We add Gaussian noise scaled to the same level of the surrounding spectrum to these linear infills to mimic a continuous spectrum. We then model and fit the broad Na doublet, as well as the nearby He\,\textsc{i} emission line, using the same methods as for our lower resolution spectra. The spectrum with the narrow line emission removed is shown in Figure \ref{fig:300otspec}.

We measure the broad components at wavelengths of 5890.013 $\pm$ 0.043 {\AA} and 5895.987 $\pm$ 0.043 {\AA}, with velocities of 268.31~$\pm$~3.38~km~s$^{-1}$ and 268.04~$\pm$~3.38~km~s$^{-1}$ respectively. We calculate an equivalent width of 1.87~$\pm$~0.01~{\AA} for the broad components of the Na doublet.

In order to demonstrate the effect of using low-resolution spectra compared to high-resolution, we degrade this spectrum to a resolution similar to the majority of our spectra using the same processes described in Section \ref{sec:synthhires}. After degrading the original UVES spectrum such that it matches the resolution of our other spectra, we again fit for the equivalent width. In this case, we find a value of 2.44~$\pm$~0.67~{\AA}. This value is consistent with that obtained from the high-resolution spectrum, though the uncertainty is much higher, illustrating the usefulness of high-resolution spectroscopy for accurate determination of equivalent widths.

As a means of comparison, we fit a number of other observed emission lines found in this spectrum which we expect to arise due to interaction with fast moving ejecta. These include the H$\alpha$ emission line and the Ca\,\textsc{ii} infrared triplet. While the H$\alpha$ line shows some evidence of substructure, we fit it as single emission line to get an estimate of its velocity. The Ca triplet is unfortunately partially cut off due to the wavelength capabilities of the UVES instrument, but the entirety of the 8500 {\AA} line is visible, and so we fit this. This emission line is accompanied by a prominent narrow absorption feature, which we include in our fit. Our fit to H$\alpha$ returns a velocity of 658.00 $\pm$ 1.45 km s$^{-1}$, while our fit to the Ca line returns a velocity of 436.27 $\pm$ 2.76 km s$^{-1}$. Within uncertainties, these velocities each exceed those determined for Na absorption. If these emission lines are produced by ejecta moving faster than the circumstellar material, this would provide evidence that the absorption we see from Na is produced in the CSM.

However, it is additionally possible that the different velocities of these lines arise not from production in different media of CSM and ejecta, rather as an effect of different optical depths. In SNe IIP, line velocities can differ by widths of 1000s of km s$^{-1}$, due to lines forming in different ejecta regions. Similarly, it may be that the difference in our velocities is also from lines being formed and observed in different parts of a dense CSM with different velocities.

\subsection{Velocity evolution}

To attempt to distinguish between Na lines arising from CSM as opposed to ejecta, we measure the evolution of the velocity of the Na doublet in our sample of ILRTs. The velocity in this section refers to the offset of the central wavelength of the D2 components of the Na line from its rest wavelength, as opposed to the width of the line. 

To do this, we first find the observed wavelength of H$\alpha$ emission in our spectra. As this line is usually strong compared to the continuum, we simply take the wavelength of H$\alpha$ to be the wavelength at the point where flux is highest within the region of the emission. We then compare this observed wavelength to the rest wavelength of H$\alpha$ and calculate a pseudo-redshift. This step is necessary due to the heterogeneity of spectra in our sample. Data from WISeREP have not necessarily been corrected for redshift from the host galaxy, and differences between observations using different instruments may introduce instrumental effects. We therefore use this pseudo-redshift as a catch-all for measuring the correction between the observed and rest spectra. Using this pseudo-redshift, we calculate the rest wavelength of the Na doublet from our fits of the observed spectra. We then compare this wavelength to the expected rest wavelength to determine the velocity of the material in which the Na doublet is being produced.

If the absorption doublet is produced in a layer of CSM, we would expect the velocity to remain relatively constant over time. In contrast, if the absorption was taking place within the ejecta, we would expect to see evolution. As the ejecta travels further from the star, its optical depth will drop, allowing us to see further into the ejecta to regions where the ejecta velocity is lower. Thus, we would expect to see the velocity of the doublet drop over time.

\begin{figure*}
    \centering
    \subfloat[]{
      \includegraphics[width=0.32\linewidth]{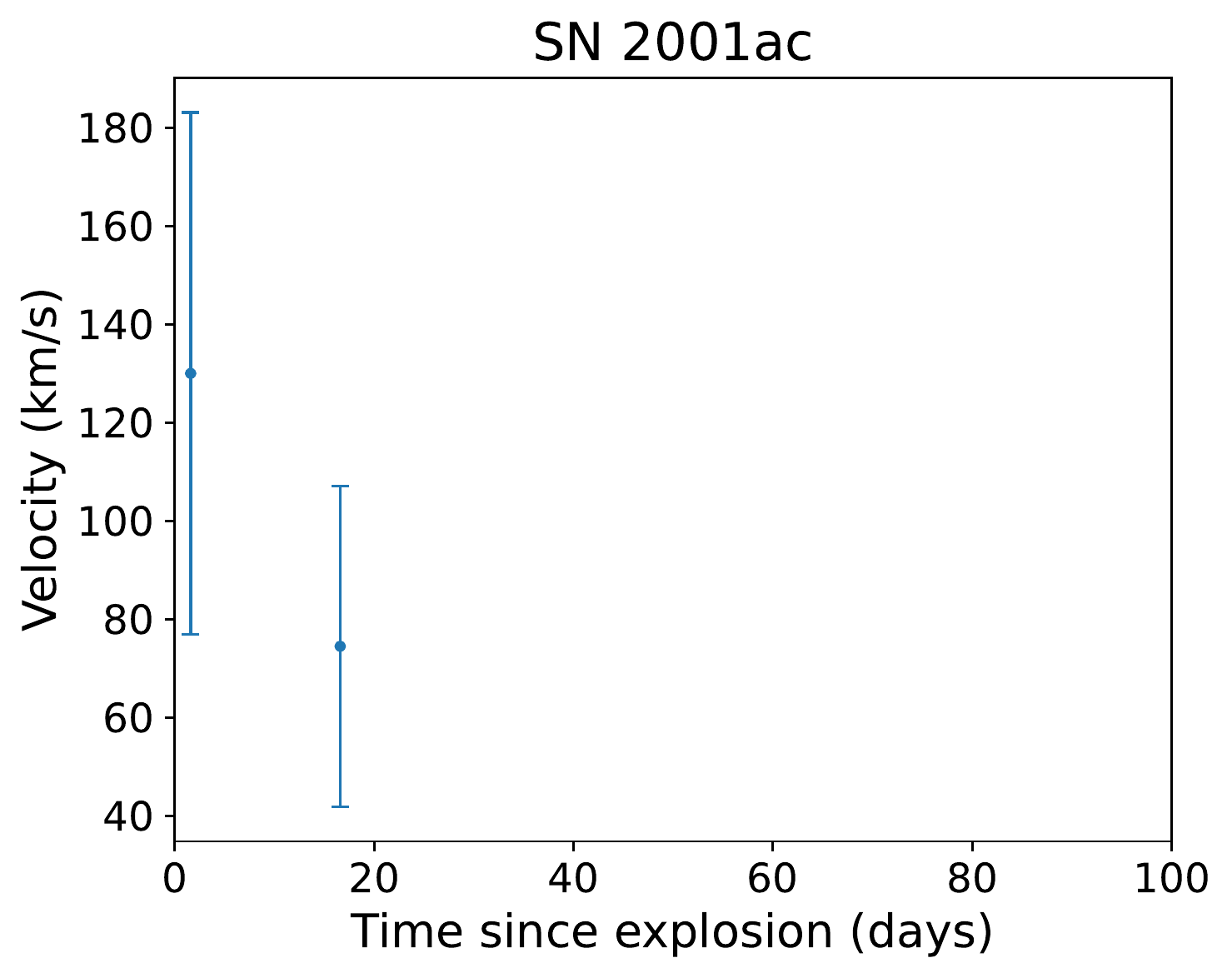}
    }
    \subfloat[]{
      \includegraphics[width=0.32\linewidth]{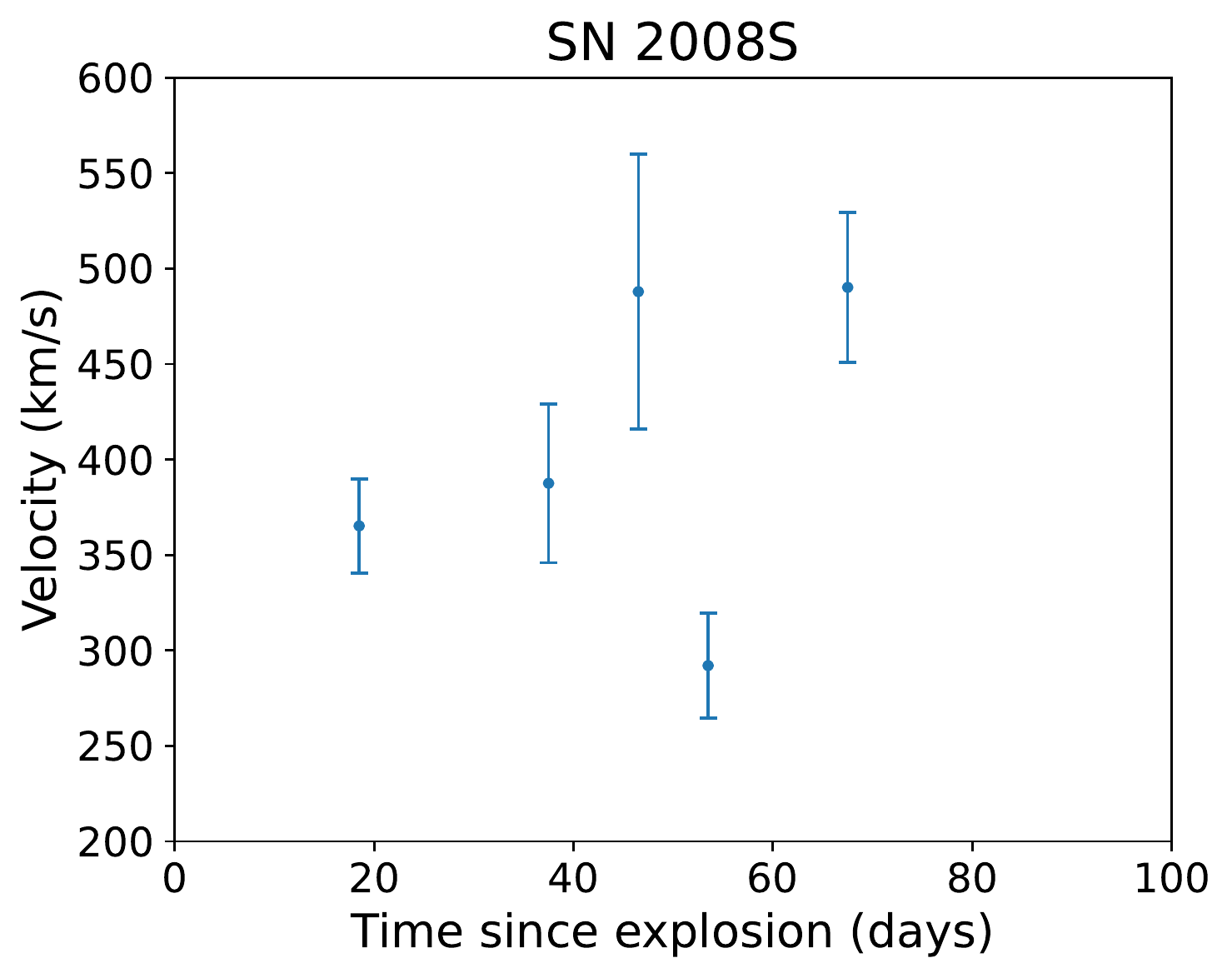}
    }
    \subfloat[]{
      \includegraphics[width=0.32\linewidth]{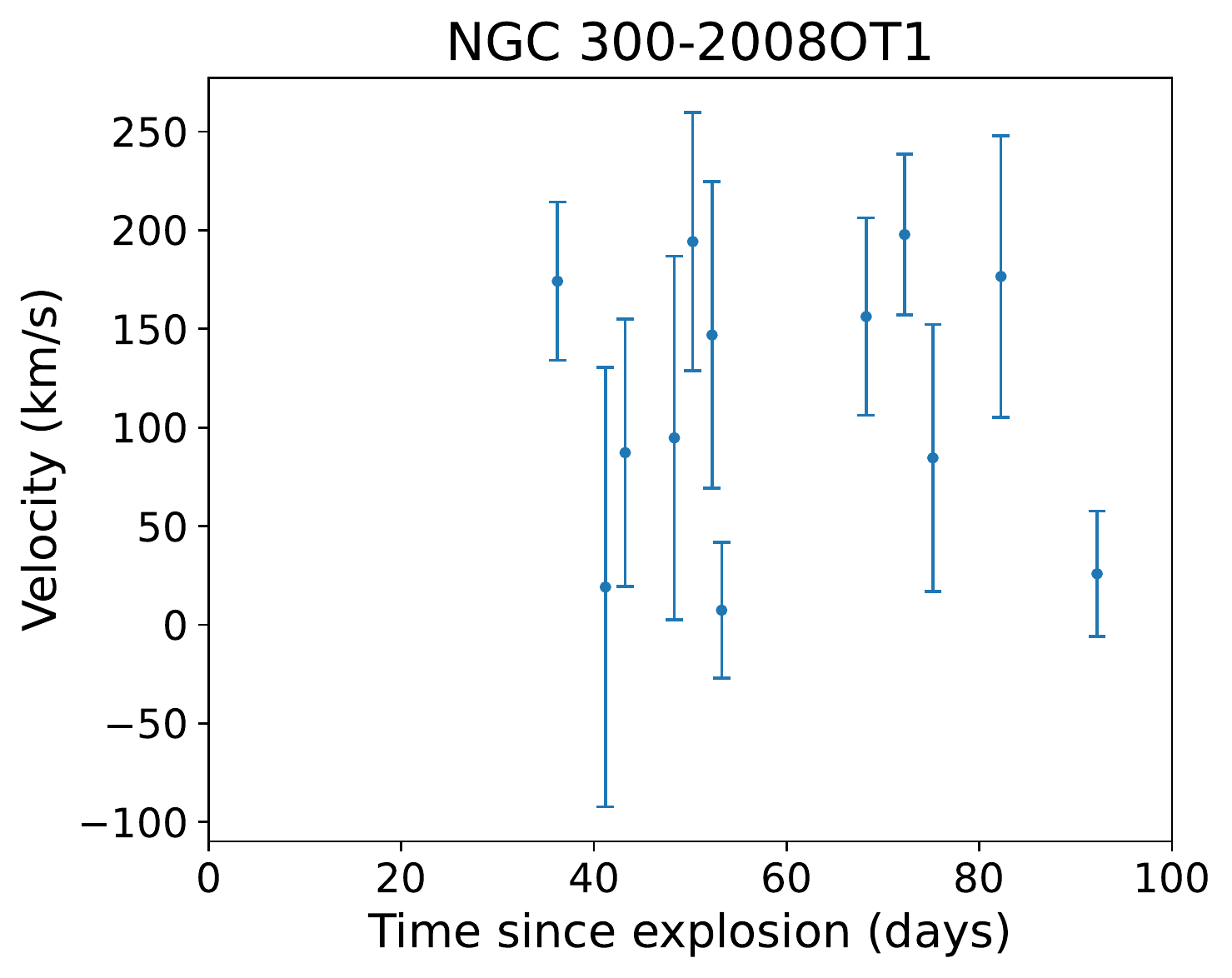}
    }
    
    \hspace{0mm}
    
    \subfloat[]{
      \includegraphics[width=0.32\linewidth]{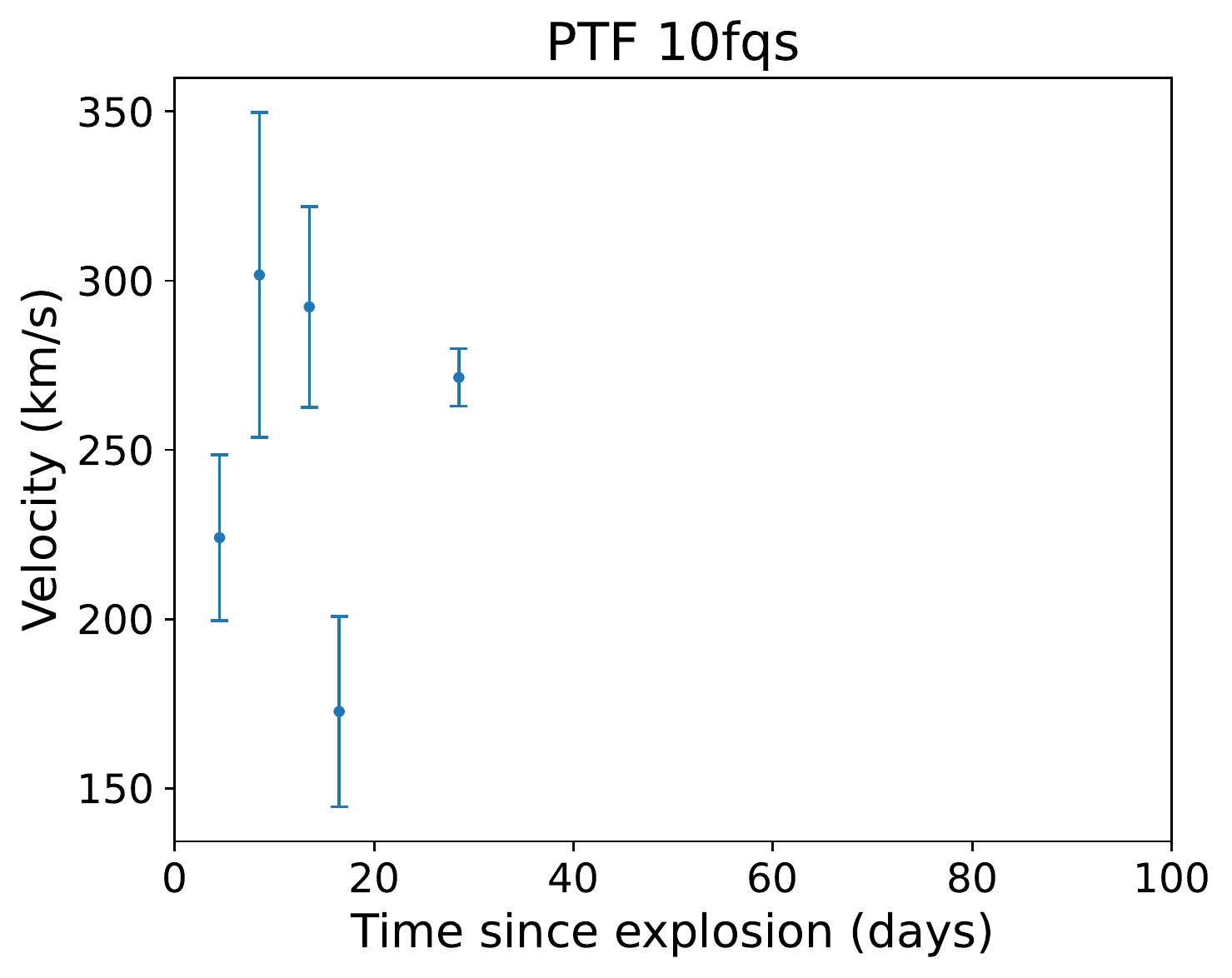}
    }
    \subfloat[]{
      \includegraphics[width=0.32\linewidth]{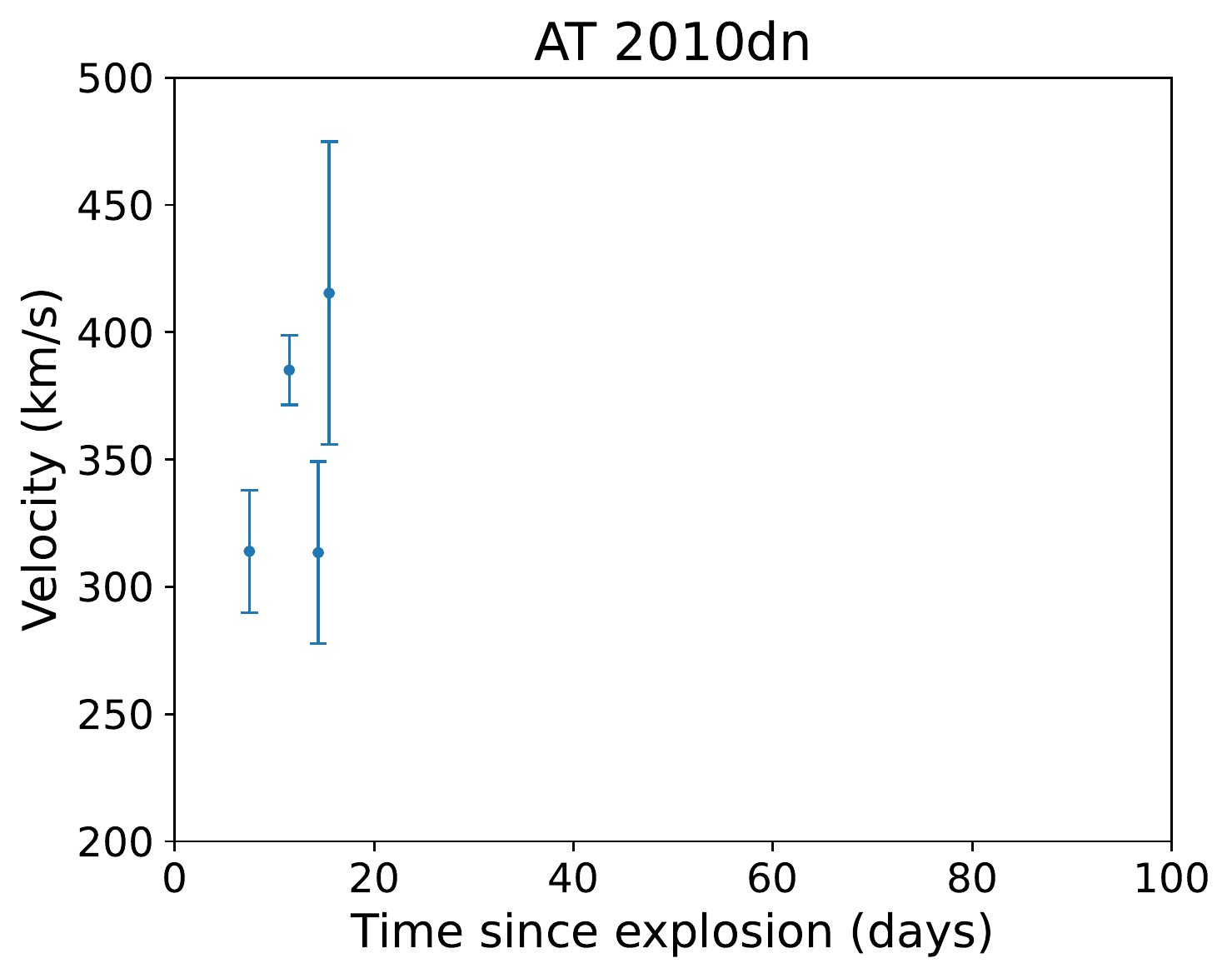}
    }
    \subfloat[]{
      \includegraphics[width=0.32\linewidth]{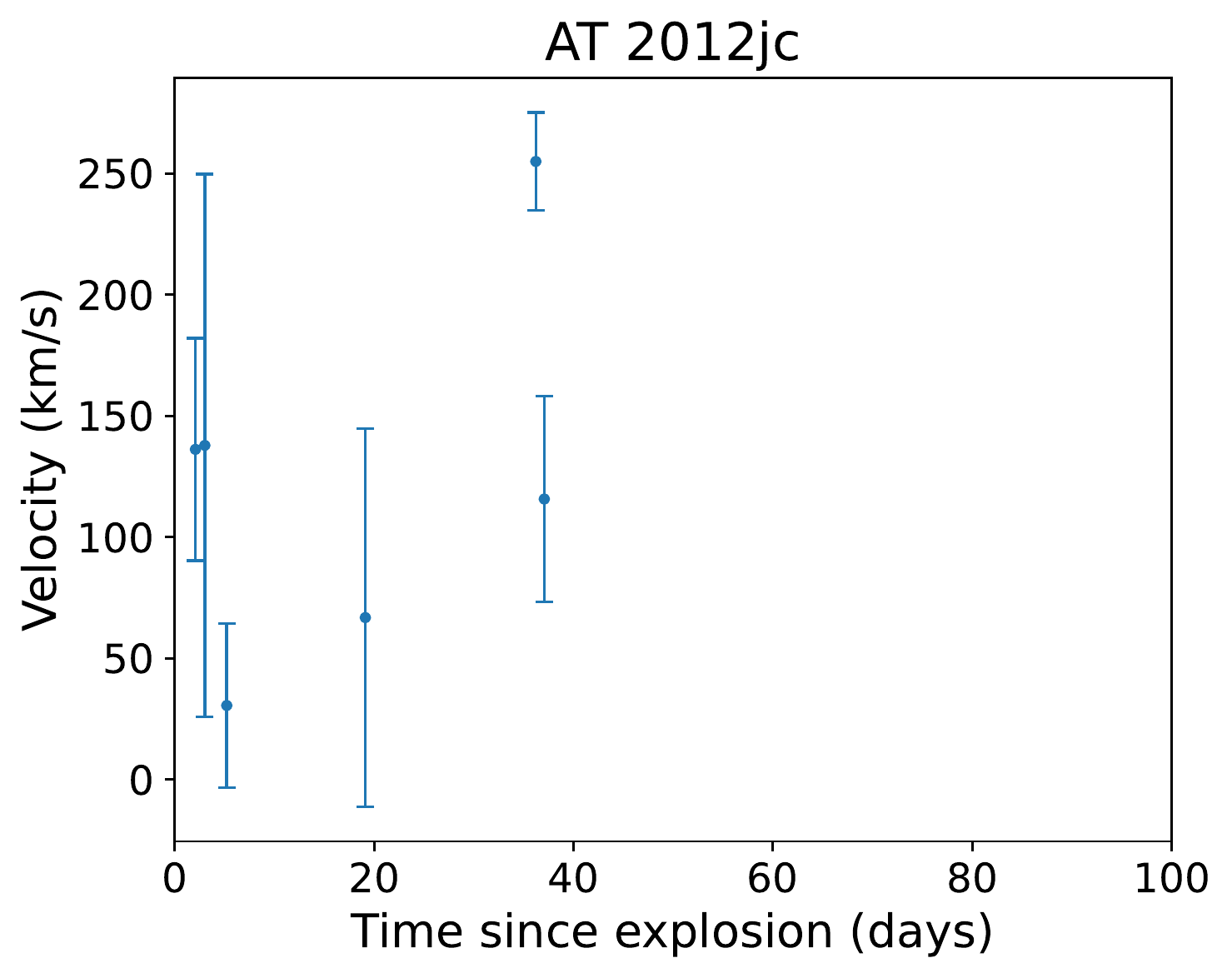}
    }
    
    \hspace{0mm}
    
    \subfloat[]{
      \includegraphics[width=0.32\linewidth]{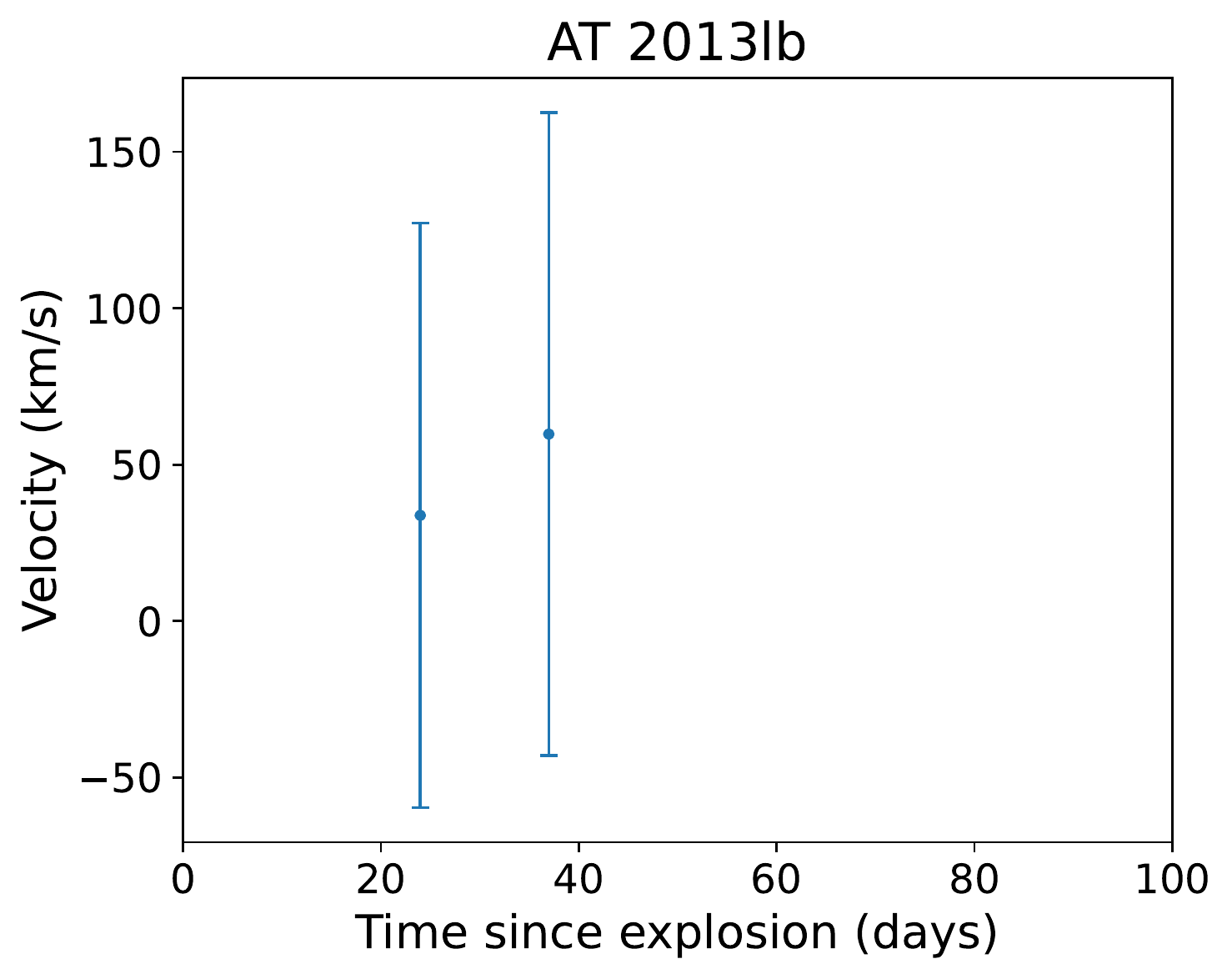}
    }
    \subfloat[]{
      \includegraphics[width=0.32\linewidth]{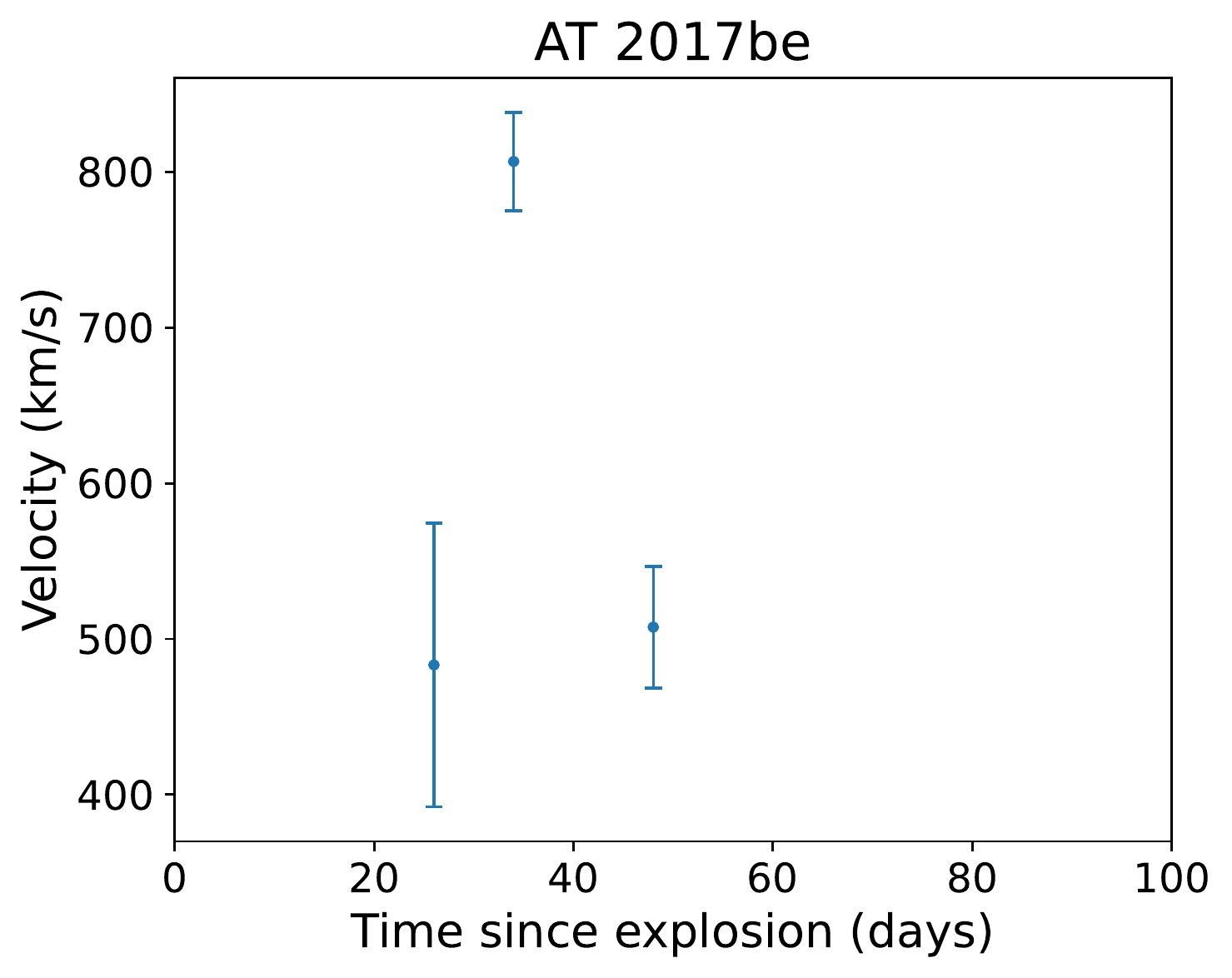}
    }
    \subfloat[]{
      \includegraphics[width=0.32\linewidth]{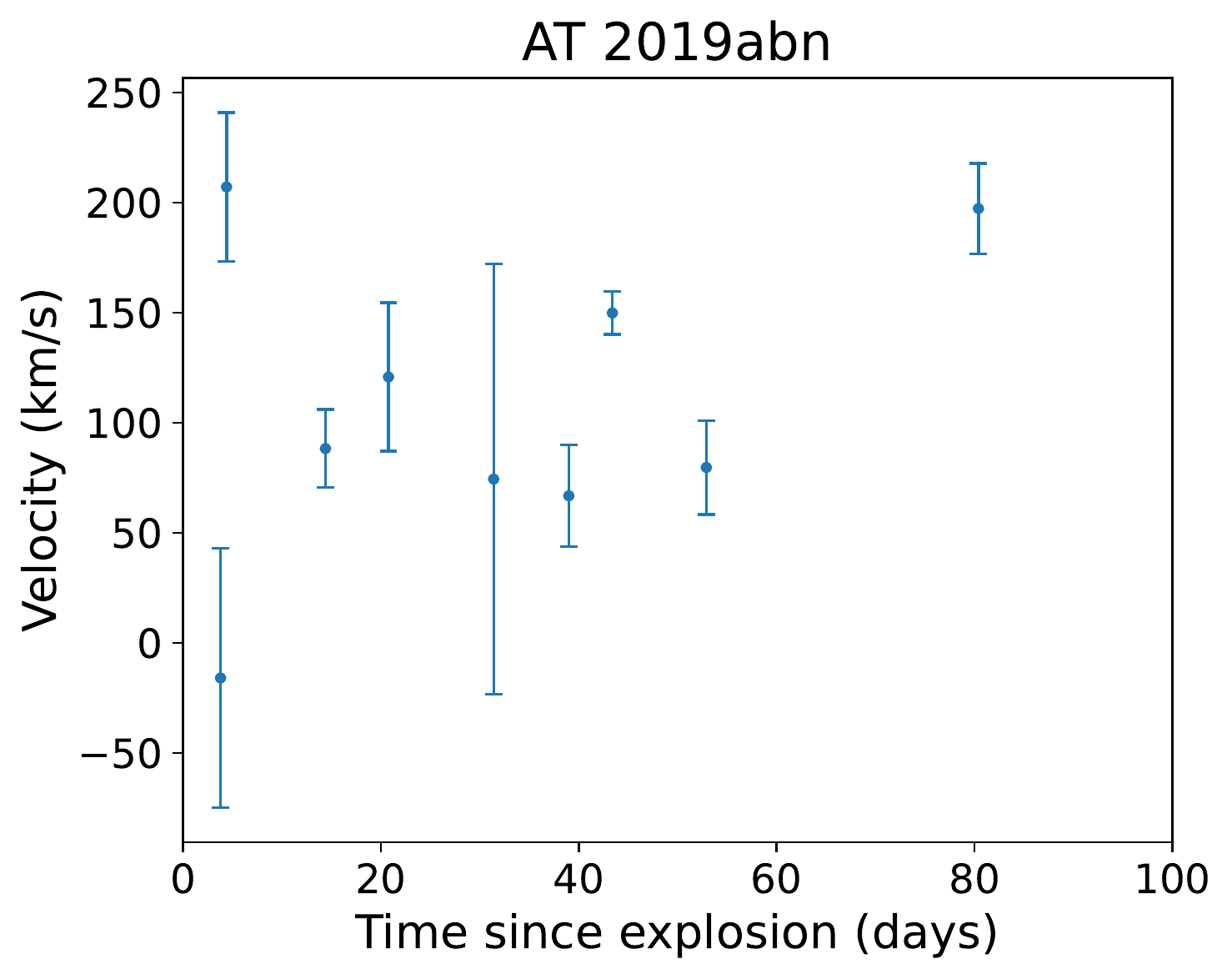}
    }

    \caption{Evolution of velocity of Na absorption doublet in our sample of ILRTs. AT 2013la not included due to lack of detections making velocity determination impossible. Uncertainties come from uncertainty of fitting central wavelengths of Na doublet, and do not factor in uncertainty in location of H$\alpha$ line used for calibration.}
    \label{fig:velocities}
\end{figure*}

Figure \ref{fig:velocities} shows the evolution of the velocity of the D2 component of the Na doublet over the first 100 days of observation for our sample of ILRTs.

The error bars plotted in Figure \ref{fig:velocities} are underestimates, including only uncertainties incurred from the fitting of the position of the Na doublet. In reality, the uncertainty in the determination of the central wavelength of H$\alpha$ would increase our uncertainties in the final calculation of velocity. Regardless, the majority of our ILRTs show little obvious evidence of velocity evolution.

Some additional caveats must be considered when interpreting these results. One is that a number of spectra from WISeREP were not usable. Several spectra, particularly for SN 2008S, showed evidence of having been shifted prior to being uploaded. In these cases, lines such as H$\alpha$ did not appear at the same observed wavelength as in other spectra, instead being 100s of {\AA} lower. Attempting to fit this discrepancy with an instrumental redshift as usual led to the position of the Na doublet being poorly estimated, resulting in velocities far higher than expected, on the order of 1000s of km s$^{-1}$. As we had no way of correcting these spectra back to their intended wavelength ranges, these spectra had to be discarded. A number of spectra for other ILRTs also showed similar behaviour.

In the other ILRTs which seem to show some evidence of evolution such as PTF 10fqs, AT 2012jc, and AT 2017be, it is worth considering that the purported evolution is only backed up by a small number of points in each case, and uncertainties are underestimated. Additionally, as issues similar to SN 2008S were present in other ILRTs, it is difficult to disentangle the true velocity of the Na line from that which may be introduced by inconsistencies between the spectra.

It may be tentatively suggested that some correlation exists between the evolutions of the velocity of the Na doublet, and its equivalent width. This is perhaps most visible in the evolutions of AT 2012jc and AT 2019abn. These purported correlations may simply arise from random chance, however we consider the implications of such a correlation, if one does indeed exist. This would imply that the strength of the absorption line is related to the velocity of material in the medium in which it is being produced. In a regime of CSM produced by a homologous wind over a large timescale, such a correlation would be impossible. In this case, material at both high and low velocities would be present in regions both close to and far from the star. On a long enough timescale, the distribution of velocities would be the same independent of distance to the star. Thus, no matter which region of CSM was being probed, the measured velocity would be the same. If instead, the CSM is produced by episodic eruptions on shorter timescales, it would be possible for a velocity gradient to exist in the CSM. In this case, higher velocity material would reside further from the star, while low velocity material would remain close. In such a case, it maybe possible that as the equivalent width of the Na line changes due to the sweeping up of material and changes in the ionisation conditions, the velocity of remaining absorbing material changes with the strength of the line.

\section{Discussions and Conclusions} \label{sec:conclusions}

We have carried out the first systematic study of the evolution of the Na\,\textsc{i} D doublet in ILRTs. The low ionisation potential of Na, combined with its relatively ubiquity in the spectra of ILRTs make it a particularly useful species for probing the nature of the CSM around these sources and thus gaining insight into the type of systems from which they arise. We present our code for fitting these spectral lines using MCMC methods and apply it to a sample of eleven ILRTs.

We find that Na\,\textsc{i} D evolution is present in our entire sample of ILRTs with a single exception where the line is not detected in any observation. This evolution displays diversity in the strength of the line, as well as its shape and time-scale. We note a number of common behaviours shared by ILRTs.

As a means of predicting the evolution displayed in ILRTs, we create a simple toy model wherein the strength of the Na\,\textsc{i} D absorption feature is directly proportional to the column density of Na\,\textsc{i} remaining in front of the ejecta from a central supernova. Testing this model using density profiles corresponding to a windy ($\propto$ r$^{-2}$) environment and a constant density environment, we predict the behaviour expected in these situations and note that two of our sample; PTF 10fqs and NGC 300-2008OT1 display evolution qualitatively similar to these.

We find that the largest issue to accurately tracking the evolution of the Na line in our sample of ILRTs was the scarcity of high quality observations available for many sources. For the majority of ILRTs in our sample our analysis relies on fewer than 10 spectra, and we only had access to a single high signal-to-noise high-resolution spectrum for one of our sample. Further observations for any of these ILRTs at both early or late times would allow us to better constrain their overall evolution and may allow us to draw parallels between other ILRTs whose similarity is only speculative with current observations.

Our toy model was successful in describing the overall shape of the Na\,\textsc{i} D evolution for two of our ILRTs, however it is clear from the diversity of evolution displayed across the whole sample that our model is too simplistic to act as a robust predictor of ILRT behaviour in general. In order to reproduce the observed evolution in greater detail, it is clear that further detailed modelling will be required. Implementation of a spectral synthesis code such as \textsc{Cloudy} \citep{cloudy} would allow for modelling with greater accuracy and the ability to reproduce some of the more complex behaviour displayed in our sample of ILRTs. This would lead to further constraints on the types of environments present in these events, and hence their associated progenitor systems.

A correlation has been reported in the literature between the strength of the Na\,\textsc{i} D line in a source and its Milky Way absorption \citep{munarizwitter, poznanskiextinction}. While this relation has often been used to estimate the extinction level for individual sources, there are a number of caveats which must be considered before using such a relation. The first is that this relation does not hold for low resolution spectra, where essentially no correlation is seen \citep{badextinction}. The second is that this correlation is an empirical trend based on a large number of spectra. While it is true that extinction increases with Na\,\textsc{i} D strength for a large population of sources, it is not necessarily true that this represents a reliable predictor for the extinction of any individual source. Finally, above a certain equivalent width, the Na\,\textsc{i} D lines begin to saturate, and the relation flattens. \cite{munarizwitter} find that this saturation occurs at equivalent widths >~0.5~{\AA} for the Na\,\textsc{i} D2 line, and while \cite{poznanskiextinction} extend this closer to 1~{\AA} by considering multiple dust clouds in the line of sight, a saturation point still remains past which the correlation becomes less useful.

For our sample of ILRTs in particular, this relation proves unhelpful. We take Equation 9 from \cite{poznanskiextinction} and naively apply it to AT 2010dn. AT 2010dn displays a maximum equivalent width of 2.88 $\pm$ 0.15 {\AA} and a minimum of 1.30 $\pm$ 0.13 {\AA}. These lead to estimates for the extinction of either $\log_{10}(E(B-V)) = -0.33 \pm 0.17$ or $\log_{10}(E(B-V)) = 1.52 \pm 0.19$, a difference of nearly two orders of magnitude. It can be seen that for a source where the strength of this line is variable in time, this relation cannot produce an accurate estimate of extinction. 

As an aside, we also note that extinction due to scattering in a spherical CSM will differ from that cause by a foreground screen. Scattering occurring in a cloud located along the line of sight to a source will serve to remove photons which would otherwise have reached us, increasing the strength of the absorption feature. Scattering from a spherical CSM introduces a second factor - while photons can still be scattered out of our line of sight, there will also be some fraction of photons which were originally travelling away from our line of sight which are instead scattered towards us through interaction with the CSM. \cite{2012aw} investigate the consequences of modelling circumstellar dust around the type IIP SN 2012aw using Galactic extinction laws, finding that ignoring spherical geometry results in overestimation of both luminosity and absorption.

The \cite{poznanskiextinction} relation is a powerful statistical tool, but we advise caution in its use as a means of predicting extinction for individual sources. Care must be taken to ensure that the Na absorption being measured is not varying in time, arises from the interstellar medium rather than CSM, is present in high-resolution spectra, and does not exceed saturation limits before it can be used as an estimate for Milky Way extinction.

The study of the Na\,\textsc{i} D doublet represents a promising new diagnostic for studying the circumstellar environment around explosive transients such as ILRTs. Further series of spectral observations will allow us to better understand these events and eventually use them as quantitative diagnostics of wind profiles.

\section*{Acknowledgements}

The research conducted in this publication was funded by the Irish Research Council under grant number GOIPG/2020/542. MF acknowledges support from a Royal Society - Science Foundation Ireland University Research Fellowship. AR acknowledges financial support from ANID BECAS/DOCTORADO NACIONAL 21202412. GV acknowledges INAF for funding his PhD fellowship within the PhD School in Astronomy at the University of Padova.

We thank Peter Duffy, Andrea Pastorello and Elena Mason for their discussions regarding this project, and the comments provided on the paper.

This work made use of Astropy:\footnote{http://www.astropy.org} a community-developed core Python package and an ecosystem of tools and resources for astronomy \citep{astropy:2013, astropy:2018, astropy:2022}

Based on observations made with ESO Telescopes at the La Silla Paranal Observatory under programme ID 281.D-5016.

\section*{Data Availability}

The spectra analysed in this study are available from the WISeREP repository at \url{https://wiserep.weizmann.ac.il}.



\bibliographystyle{mnras}
\bibliography{example} 

\begin{thebibliography}{}
\makeatletter
\relax
\def\mn@urlcharsother{\let\do\@makeother \do\$\do\&\do\#\do\^\do\_\do\%\do\~}
\def\mn@doi{\begingroup\mn@urlcharsother \@ifnextchar [ {\mn@doi@}
  {\mn@doi@[]}}
\def\mn@doi@[#1]#2{\def\@tempa{#1}\ifx\@tempa\@empty \href
  {http://dx.doi.org/#2} {doi:#2}\else \href {http://dx.doi.org/#2} {#1}\fi
  \endgroup}
\def\mn@eprint#1#2{\mn@eprint@#1:#2::\@nil}
\def\mn@eprint@arXiv#1{\href {http://arxiv.org/abs/#1} {{\tt arXiv:#1}}}
\def\mn@eprint@dblp#1{\href {http://dblp.uni-trier.de/rec/bibtex/#1.xml}
  {dblp:#1}}
\def\mn@eprint@#1:#2:#3:#4\@nil{\def\@tempa {#1}\def\@tempb {#2}\def\@tempc
  {#3}\ifx \@tempc \@empty \let \@tempc \@tempb \let \@tempb \@tempa \fi \ifx
  \@tempb \@empty \def\@tempb {arXiv}\fi \@ifundefined
  {mn@eprint@\@tempb}{\@tempb:\@tempc}{\expandafter \expandafter \csname
  mn@eprint@\@tempb\endcsname \expandafter{\@tempc}}}

\bibitem[\protect\citeauthoryear{Adams, Kochanek, Prieto, Dai, Shappee  \&
  Stanek}{Adams et~al.}{2016}]{adams300ot}
Adams S.~M.,  Kochanek C.~S.,  Prieto J.~L.,  Dai X.,  Shappee B.~J.,   Stanek
  K.~Z.,  2016, \mn@doi [Monthly Notices of the Royal Astronomical Society]
  {10.1093/mnras/stw1059}, 460, 1645

\bibitem[\protect\citeauthoryear{{Astropy Collaboration} et~al.,}{{Astropy
  Collaboration} et~al.}{2013}]{astropy:2013}
{Astropy Collaboration} et~al., 2013, \mn@doi [\aap]
  {10.1051/0004-6361/201322068}, \href
  {http://adsabs.harvard.edu/abs/2013A%26A...558A..33A} {558, A33}

\bibitem[\protect\citeauthoryear{{Astropy Collaboration} et~al.,}{{Astropy
  Collaboration} et~al.}{2018}]{astropy:2018}
{Astropy Collaboration} et~al., 2018, \mn@doi [\aj] {10.3847/1538-3881/aabc4f},
  \href {https://ui.adsabs.harvard.edu/abs/2018AJ....156..123A} {156, 123}

\bibitem[\protect\citeauthoryear{{Astropy Collaboration} et~al.,}{{Astropy
  Collaboration} et~al.}{2022}]{astropy:2022}
{Astropy Collaboration} et~al., 2022, \mn@doi [apj] {10.3847/1538-4357/ac7c74},
  \href {https://ui.adsabs.harvard.edu/abs/2022ApJ...935..167A} {935, 167}

\bibitem[\protect\citeauthoryear{{Berger} et~al.,}{{Berger}
  et~al.}{2009}]{berger}
{Berger} E.,  et~al., 2009, \mn@doi [\apj] {10.1088/0004-637X/699/2/1850},
  \href {https://ui.adsabs.harvard.edu/abs/2009ApJ...699.1850B} {699, 1850}

\bibitem[\protect\citeauthoryear{Blondin, Prieto, Patat, Challis, Hicken,
  Kirshner, Matheson  \& Modjaz}{Blondin et~al.}{2009}]{Ianovar}
Blondin S.,  Prieto J.~L.,  Patat F.,  Challis P.,  Hicken M.,  Kirshner R.~P.,
   Matheson T.,   Modjaz M.,  2009, \mn@doi [The Astrophysical Journal]
  {10.1088/0004-637x/693/1/207}, 693, 207

\bibitem[\protect\citeauthoryear{Bond, Bedin, Bonanos, Humphreys, Monard,
  Prieto  \& Walter}{Bond et~al.}{2009}]{bond300ot}
Bond H.~E.,  Bedin L.~R.,  Bonanos A.~Z.,  Humphreys R.~M.,  Monard L. A.
  G.~B.,  Prieto J.~L.,   Walter F.~M.,  2009, \mn@doi [The Astrophysical
  Journal] {10.1088/0004-637x/695/2/l154}, 695, L154

\bibitem[\protect\citeauthoryear{{Borkowski}, {Blondin}  \&
  {Reynolds}}{{Borkowski} et~al.}{2009}]{borkowski}
{Borkowski} K.~J.,  {Blondin} J.~M.,   {Reynolds} S.~P.,  2009, \mn@doi [\apjl]
  {10.1088/0004-637X/699/2/L64}, \href
  {https://ui.adsabs.harvard.edu/abs/2009ApJ...699L..64B} {699, L64}

\bibitem[\protect\citeauthoryear{Botticella et~al.,}{Botticella
  et~al.}{2009}]{botticella08s}
Botticella M.~T.,  et~al., 2009, \mn@doi [Monthly Notices of the Royal
  Astronomical Society] {10.1111/j.1365-2966.2009.15082.x}, 398, 1041

\bibitem[\protect\citeauthoryear{{Bowen}, {Roth}, {Meyer}  \& {Blades}}{{Bowen}
  et~al.}{2000}]{bowen}
{Bowen} D.~V.,  {Roth} K.~C.,  {Meyer} D.~M.,   {Blades} J.~C.,  2000, \mn@doi
  [\apj] {10.1086/308913}, \href
  {https://ui.adsabs.harvard.edu/abs/2000ApJ...536..225B} {536, 225}

\bibitem[\protect\citeauthoryear{{Brown} et~al.,}{{Brown}
  et~al.}{2019}]{2017erp}
{Brown} P.~J.,  et~al., 2019, \mn@doi [\apj] {10.3847/1538-4357/ab1a3f}, \href
  {https://ui.adsabs.harvard.edu/abs/2019ApJ...877..152B} {877, 152}

\bibitem[\protect\citeauthoryear{Cai et~al.,}{Cai et~al.}{2018}]{cai17be}
Cai Y.-Z.,  et~al., 2018, \mn@doi [Monthly Notices of the Royal Astronomical
  Society] {10.1093/mnras/sty2070}

\bibitem[\protect\citeauthoryear{Cai et~al.,}{Cai et~al.}{2021}]{caiILRTs}
Cai Y.-Z.,  et~al., 2021, \mn@doi [Astronomy {\&} Astrophysics]
  {10.1051/0004-6361/202141078}, 654, A157

\bibitem[\protect\citeauthoryear{{Cai}, {Reguitti}, {Valerin}  \& {Wang}}{{Cai}
  et~al.}{2022a}]{caigap}
{Cai} Y.,  {Reguitti} A.,  {Valerin} G.,   {Wang} X.,  2022a, \mn@doi
  [Universe] {10.3390/universe8100493}, \href
  {https://ui.adsabs.harvard.edu/abs/2022Univ....8..493C} {8, 493}

\bibitem[\protect\citeauthoryear{{Cai} et~al.,}{{Cai} et~al.}{2022b}]{21biy}
{Cai} Y.~Z.,  et~al., 2022b, \mn@doi [\aap] {10.1051/0004-6361/202244393},
  \href {https://ui.adsabs.harvard.edu/abs/2022A&A...667A...4C} {667, A4}

\bibitem[\protect\citeauthoryear{{Chugai} \& {Utrobin}}{{Chugai} \&
  {Utrobin}}{2008}]{chugai}
{Chugai} N.~N.,  {Utrobin} V.~P.,  2008, \mn@doi [Astronomy Letters]
  {10.1134/S1063773708090028}, \href
  {https://ui.adsabs.harvard.edu/abs/2008AstL...34..589C} {34, 589}

\bibitem[\protect\citeauthoryear{Doherty, Gil-Pons, Siess  \&
  Lattanzio}{Doherty et~al.}{2017}]{dohertySAGB}
Doherty C.~L.,  Gil-Pons P.,  Siess L.,   Lattanzio J.~C.,  2017, \mn@doi
  [Publications of the Astronomical Society of Australia]
  {10.1017/pasa.2017.52}, 34

\bibitem[\protect\citeauthoryear{{Ferland} et~al.,}{{Ferland}
  et~al.}{2017}]{cloudy}
{Ferland} G.~J.,  et~al., 2017, \rmxaa, \href
  {https://ui.adsabs.harvard.edu/abs/2017RMxAA..53..385F} {53, 385}

\bibitem[\protect\citeauthoryear{{Foreman-Mackey}, {Hogg}, {Lang}  \&
  {Goodman}}{{Foreman-Mackey} et~al.}{2013}]{emcee}
{Foreman-Mackey} D.,  {Hogg} D.~W.,  {Lang} D.,   {Goodman} J.,  2013, \mn@doi
  [\pasp] {10.1086/670067}, \href
  {https://ui.adsabs.harvard.edu/abs/2013PASP..125..306F} {125, 306}

\bibitem[\protect\citeauthoryear{{Humphreys}, {Bond}, {Bedin}, {Bonanos},
  {Davidson}, {Berto Monard}, {Prieto}  \& {Walter}}{{Humphreys}
  et~al.}{2011}]{humphreys300ot}
{Humphreys} R.~M.,  {Bond} H.~E.,  {Bedin} L.~R.,  {Bonanos} A.~Z.,  {Davidson}
  K.,  {Berto Monard} L.~A.~G.,  {Prieto} J.~L.,   {Walter} F.~M.,  2011,
  \mn@doi [\apj] {10.1088/0004-637X/743/2/118}, \href
  {https://ui.adsabs.harvard.edu/abs/2011ApJ...743..118H} {743, 118}

\bibitem[\protect\citeauthoryear{Jencson et~al.,}{Jencson
  et~al.}{2019}]{jencson19abn}
Jencson J.~E.,  et~al., 2019, \mn@doi [The Astrophysical Journal]
  {10.3847/2041-8213/ab2c05}, 880, L20

\bibitem[\protect\citeauthoryear{Kasliwal et~al.,}{Kasliwal
  et~al.}{2011}]{kasliwalLRNe}
Kasliwal M.~M.,  et~al., 2011, \mn@doi [The Astrophysical Journal]
  {10.1088/0004-637x/730/2/134}, 730, 134

\bibitem[\protect\citeauthoryear{{Kochanek}, {Khan}  \& {Dai}}{{Kochanek}
  et~al.}{2012}]{2012aw}
{Kochanek} C.~S.,  {Khan} R.,   {Dai} X.,  2012, \mn@doi [\apj]
  {10.1088/0004-637X/759/1/20}, \href
  {https://ui.adsabs.harvard.edu/abs/2012ApJ...759...20K} {759, 20}

\bibitem[\protect\citeauthoryear{{Kulkarni} \& {Kasliwal}}{{Kulkarni} \&
  {Kasliwal}}{2009}]{kulkarnigap}
{Kulkarni} S.,  {Kasliwal} M.~M.,  2009, in {Kawai} N.,  {Mihara} T.,  {Kohama}
  M.,   {Suzuki} M.,  eds, Astrophysics with All-Sky X-Ray Observations. p.~312
  (\mn@eprint {arXiv} {0903.0218})

\bibitem[\protect\citeauthoryear{{Mackay}}{{Mackay}}{2003}]{MCMC}
{Mackay} D. J.~C.,  2003, {Information Theory, Inference and Learning
  Algorithms}

\bibitem[\protect\citeauthoryear{{Maguire} et~al.,}{{Maguire}
  et~al.}{2013}]{Maguire2013}
{Maguire} K.,  et~al., 2013, \mn@doi [\mnras] {10.1093/mnras/stt1586}, \href
  {https://ui.adsabs.harvard.edu/abs/2013MNRAS.436..222M} {436, 222}

\bibitem[\protect\citeauthoryear{{Munari} \& {Zwitter}}{{Munari} \&
  {Zwitter}}{1997}]{munarizwitter}
{Munari} U.,  {Zwitter} T.,  1997, \aap, \href
  {https://ui.adsabs.harvard.edu/abs/1997A&A...318..269M} {318, 269}

\bibitem[\protect\citeauthoryear{Pastorello \& Fraser}{Pastorello \&
  Fraser}{2019}]{pastorellogap}
Pastorello A.,  Fraser M.,  2019, \mn@doi [Nature Astronomy]
  {10.1038/s41550-019-0809-9}, 3, 676

\bibitem[\protect\citeauthoryear{{Pastorello} et~al.,}{{Pastorello}
  et~al.}{2009}]{2005cs}
{Pastorello} A.,  et~al., 2009, \mn@doi [\mnras]
  {10.1111/j.1365-2966.2009.14505.x}, \href
  {https://ui.adsabs.harvard.edu/abs/2009MNRAS.394.2266P} {394, 2266}

\bibitem[\protect\citeauthoryear{Patat et~al.,}{Patat et~al.}{2007}]{patat06x}
Patat F.,  et~al., 2007, \mn@doi [Science] {10.1126/science.1143005}, 317, 924

\bibitem[\protect\citeauthoryear{{Poznanski}, {Ganeshalingam}, {Silverman}  \&
  {Filippenko}}{{Poznanski} et~al.}{2011}]{badextinction}
{Poznanski} D.,  {Ganeshalingam} M.,  {Silverman} J.~M.,   {Filippenko} A.~V.,
  2011, \mn@doi [\mnras] {10.1111/j.1745-3933.2011.01084.x}, \href
  {https://ui.adsabs.harvard.edu/abs/2011MNRAS.415L..81P} {415, L81}

\bibitem[\protect\citeauthoryear{{Poznanski}, {Prochaska}  \&
  {Bloom}}{{Poznanski} et~al.}{2012}]{poznanskiextinction}
{Poznanski} D.,  {Prochaska} J.~X.,   {Bloom} J.~S.,  2012, \mn@doi [\mnras]
  {10.1111/j.1365-2966.2012.21796.x}, \href
  {https://ui.adsabs.harvard.edu/abs/2012MNRAS.426.1465P} {426, 1465}

\bibitem[\protect\citeauthoryear{Prieto et~al.,}{Prieto
  et~al.}{2008}]{prieto08sprogen}
Prieto J.~L.,  et~al., 2008, \mn@doi [The Astrophysical Journal]
  {10.1086/589922}, 681, L9

\bibitem[\protect\citeauthoryear{Risberg}{Risberg}{1956}]{NaD}
Risberg P.,  1956, Ark. Fys., 10, 583

\bibitem[\protect\citeauthoryear{{STScI Development Team}}{{STScI Development
  Team}}{2013}]{pysynphot}
{STScI Development Team} 2013, {pysynphot: Synthetic photometry software
  package} (\mn@eprint {ascl} {1303.023})

\bibitem[\protect\citeauthoryear{Savitzky \& Golay}{Savitzky \&
  Golay}{1964}]{savgol}
Savitzky A.,  Golay M. J.~E.,  1964, \mn@doi [Analytical Chemistry]
  {10.1021/ac60214a047}, 36, 1627

\bibitem[\protect\citeauthoryear{{Shivvers} et~al.,}{{Shivvers}
  et~al.}{2017}]{2017PASP..129e4201S}
{Shivvers} I.,  et~al., 2017, \mn@doi [\pasp] {10.1088/1538-3873/aa54a6}, \href
  {https://ui.adsabs.harvard.edu/abs/2017PASP..129e4201S} {129, 054201}

\bibitem[\protect\citeauthoryear{Shore, Augusteijn, Ederoclite  \& Uthas}{Shore
  et~al.}{2011}]{shoreTPyx}
Shore S.~N.,  Augusteijn T.,  Ederoclite A.,   Uthas H.,  2011, \mn@doi
  [Astronomy {\&} Astrophysics] {10.1051/0004-6361/201117721}, 533, L8

\bibitem[\protect\citeauthoryear{Smith et~al.,}{Smith et~al.}{2009}]{smithLBVs}
Smith N.,  et~al., 2009, \mn@doi [The Astrophysical Journal]
  {10.1088/0004-637x/697/1/l49}, 697, L49

\bibitem[\protect\citeauthoryear{{Smith}, {Li}, {Silverman}, {Ganeshalingam}
  \& {Filippenko}}{{Smith} et~al.}{2011}]{2011MNRAS.415..773S}
{Smith} N.,  {Li} W.,  {Silverman} J.~M.,  {Ganeshalingam} M.,   {Filippenko}
  A.~V.,  2011, \mn@doi [\mnras] {10.1111/j.1365-2966.2011.18763.x}, \href
  {https://ui.adsabs.harvard.edu/abs/2011MNRAS.415..773S} {415, 773}

\bibitem[\protect\citeauthoryear{{Soker}}{{Soker}}{2014}]{soker}
{Soker} N.,  2014, \mn@doi [\mnras] {10.1093/mnrasl/slu119}, \href
  {https://ui.adsabs.harvard.edu/abs/2014MNRAS.444L..73S} {444, L73}

\bibitem[\protect\citeauthoryear{{Sternberg} et~al.,}{{Sternberg}
  et~al.}{2011}]{Sternberg2011}
{Sternberg} A.,  et~al., 2011, \mn@doi [Science] {10.1126/science.1203836},
  \href {https://ui.adsabs.harvard.edu/abs/2011Sci...333..856S} {333, 856}

\bibitem[\protect\citeauthoryear{{Stritzinger} et~al.,}{{Stritzinger}
  et~al.}{2020}]{2012jc}
{Stritzinger} M.~D.,  et~al., 2020, \mn@doi [\aap]
  {10.1051/0004-6361/202038018}, \href
  {https://ui.adsabs.harvard.edu/abs/2020A&A...639A.103S} {639, A103}

\bibitem[\protect\citeauthoryear{{Tody}}{{Tody}}{1986}]{iraf}
{Tody} D.,  1986, in {Crawford} D.~L.,  ed.,  Society of Photo-Optical
  Instrumentation Engineers (SPIE) Conference Series Vol. 627, Instrumentation
  in astronomy VI. p.~733, \mn@doi{10.1117/12.968154}

\bibitem[\protect\citeauthoryear{Turatto, Benetti  \& Cappellaro}{Turatto
  et~al.}{2003}]{variety}
Turatto M.,  Benetti S.,   Cappellaro E.,  2003, in Hillebrandt W.,  Leibundgut
  B.,  eds, From Twilight to Highlight: The Physics of Supernovae. Springer
  Berlin Heidelberg, Berlin, Heidelberg, pp 200--209

\bibitem[\protect\citeauthoryear{{Yaron} \& {Gal-Yam}}{{Yaron} \&
  {Gal-Yam}}{2012}]{wiserep}
{Yaron} O.,  {Gal-Yam} A.,  2012, \mn@doi [\pasp] {10.1086/666656}, \href
  {https://ui.adsabs.harvard.edu/abs/2012PASP..124..668Y} {124, 668}

\bibitem[\protect\citeauthoryear{{de Jaeger} et~al.,}{{de Jaeger}
  et~al.}{2015}]{dejaeger11a}
{de Jaeger} T.,  et~al., 2015, \mn@doi [\apj] {10.1088/0004-637X/807/1/63},
  \href {https://ui.adsabs.harvard.edu/abs/2015ApJ...807...63D} {807, 63}

\makeatother
\end{thebibliography}




\appendix

\section{Line Fitting} \label{app:line}

This appendix provides a detailed description of the code which we use to analyse each of the spectra in our sample of ILRTs. 

\begin{figure}
    \centering
    \includegraphics[width = \linewidth]{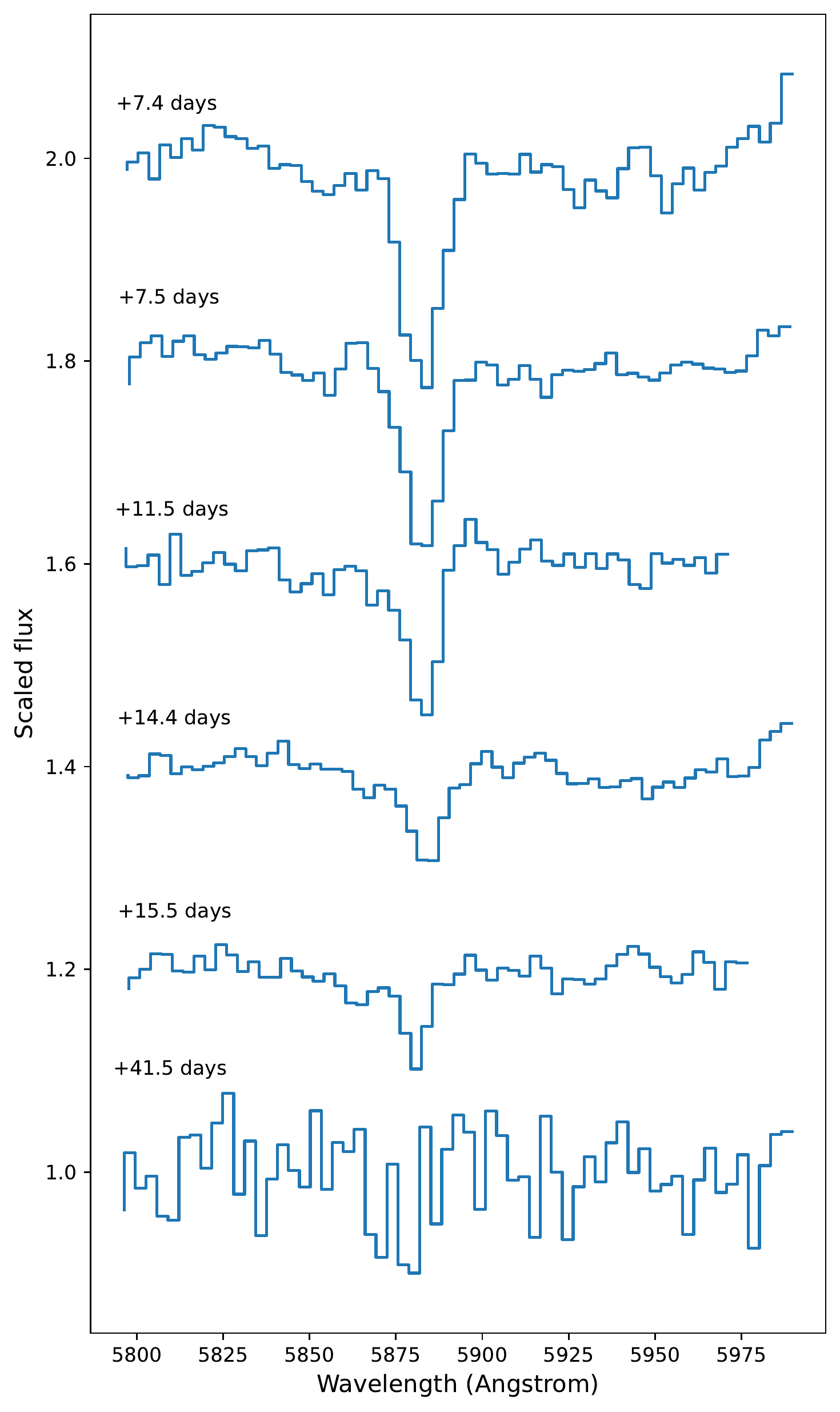}
    \caption{Spectral evolution of the Na\,\textsc{i} D line for the ILRT AT 2010dn. Spectrum has been trimmed to the area around the absorption feature, and continuum has been normalised. It can be seen that the strength of the absorption line with respect to the continuum weakens over time. Each spectrum has been vertically offset by a value of 0.2.}
    \label{fig:specsequence}
\end{figure}

We begin by trimming each spectrum down to a region centred on the expected wavelength of Na\,\textsc{i} D absorption scaled to the redshift of the host galaxy. The total width of this region was 200 {\AA}. We found that for each of our spectra this included enough of the continuum to make a satisfactory fit. We normalise the spectra such that they have a flat continuum at 1. This is performed by first manually selecting regions of continuum, excluding the central Na absorption or other nearby lines. A polynomial of user-specified order is then applied to the chosen continuum, and the spectrum is divided by this fit, flattening the continuum to unity. Figure \ref{fig:specsequence} shows an example sequence of spectra after these initial processing steps are taken.

Our fitting requires an initial estimate on the uncertainties associated with each flux measurement in our spectra. As many of the spectra available on WISeREP do not include uncertainties, we estimate our own. We apply a Savitzky-Golay filter \citep{savgol} to each spectrum with a filter window length of 11, fitting fifth order polynomials. We find that at the spectral resolutions available in these spectra, these parameters smooth out the high frequency noise while retaining much of the information from broader features such as the Na absorption doublet, producing a smoothed version of the spectrum. At each point, the difference between the smoothed and original spectrum is calculated, and the standard deviation of these differences is calculated. The uncertainty for each normalised flux measurement is taken as this value multiplied by the square root of the normalised flux at that point, to capture its Poissonian variance.

The Na\,\textsc{i} D doublet is fit for each spectrum via the implementation of MCMC methods.

We model each spectrum individually, beginning with a basic model for a linear continuum defined by two free variables. While we normalise our spectra such that the slope and y-intercept of the continuum should equal 0 and 1 respectively, we allow the slope and intercept of the continuum that we fit to vary in order to quantify their associated uncertainties. The Na doublet is modelled as two Gaussians using a total of four free variables. The amplitude of each Gaussian is set as a variable. As both lines are produced in the same material we expect their velocities, and hence widths to be equal, and so we fit for a single width which is applied to both Gaussians. Finally, we expect the separation between the two lines of the doublet to be given by the intrinsic difference between the wavelengths of each component ($\sim$~6 {\AA}). Therefore, we fit for only the central wavelength of one of the Gaussians, and fix the position of the second relative to this.

One assumption being made is that each of the lines in the doublet can be accurately fit by symmetrical Gaussians. In reality, relative motions including that of the CSM around the central supernova and that of the star in its host galaxy may cause the line profiles to show asymmetry. We see small levels of asymmetry in some of our spectra, however we choose not to fit asymmetrical Gaussians as the effect of this asymmetry is small and would lead to increases in the complexity of the model and computation time which would not be commensurate to the increase in accuracy of fitting the equivalent width.

In the case of one source, separate Na absorption features can be distinguished at redshifts for the Milky Way as well as for the host galaxy. In this case, we fit both Na doublets simultaneously, specifying an initial guess for the wavelengths of each instance. For many ILRTs in our sample, the host galaxy redshift is low enough that absorption from the host and from the Milky Way is blended. In these cases, we simply fit a single Na doublet. While these measurements will include absorption from within the Milky Way, this absorption should remain constant throughout observations, and therefore should not affect the evolution displayed by the ILRT in question. In the cases where separation exists between the host and Milky Way Na absorption we found that only contribution from the host was visible, with the one aforementioned exception.

One parameter in our model is used to account for under- or over-estimation of the uncertainties in our spectra. As many of our spectra do not contain uncertainties we estimate them instead, and this parameter accounts for the difference between our estimates and true values for uncertainty. We implement this by choosing our likelihood function to be a Gaussian where the variance is underestimated by some fractional amount $f$. We then fit for the logarithm of $f$ in our MCMC model to determine the true amount of over- or under-estimation present in our initial guesses for uncertainty.

In a number of cases, nearby spectral lines are visible, particularly He\,\textsc{i} emission at $\sim$ 5875 {\AA}. In these cases, modelling the continuum as flat is insufficient, and so we allow for the fitting of additional Gaussian components unrelated to the Na absorption doublet. In each spectrum where this occurs, we manually specify the approximate wavelengths of any additional emission or absorption features we wish to fit in our spectrum, and our model adds these to the fit.

We begin with initial guesses for each variable in the model. For our continuum, we start by assuming that the slope is 0 and the intercept is 1. 

For Na doublets, we initialise the central wavelength of the first Gaussian as the known value for the D2 component scaled to the redshift of the host galaxy. In cases where we also fit the Milky Way absorption, we initialise a second doublet with a central wavelength for the first Gaussian given as the known rest wavelength for the D2 component (5889.95 {\AA}). We initialise the width of both Gaussians in each to be 3 {\AA}.

We initialise the amplitude of the D2 component of the Na doublet to equal one minus the minimum value of flux present in the trimmed spectrum as the Na absorption is typically the strongest absorption feature in this subregion of the spectrum. We then initialise the amplitude of the D1 component to be half this value, as the D2 component is expected to be roughly twice as intense as the D1 component \citep{NaD}. 

For any further emission or absorption lines in our model, we initialise their central wavelengths at the user-specified positions, their widths as 3 {\AA}, and their amplitudes as either the flux at this position minus one for emission lines, or one minus the flux at this position for absorption lines.

We initialise 30 walkers using the \textsc{emcee} code, and begin them at values perturbed from the initial guesses up to a value of $10^{-5}$ for each parameter. We set each of these walkers to carry out a walk of 2500 steps to produce individual Markov chains. We tested various configurations of these parameters, and found that these returned satisfactory fits in a reasonable amount of computing time. Further information on the accuracy of these fits is available in Section \ref{sec:synthhires}.

We attempt to minimise the $\chi^2$ value between the model defined by the walker at each step and the true values of the spectrum. This $\chi^2$ term includes the contribution from the $\log{f}$ term introduced to account for under- or over-estimations of our uncertainties. We impose boundaries for each of the parameters such that any steps which cause a parameter to leave these boundaries returns a probability of $-\infty$, ensuring that our results stay within these bounds.

We set our bounds for wavelength of the D2 component of the Na doublet to be within 5 {\AA} of the specified central wavelength. The amplitude of this component is constrained between 0 and 1.5 times our initial guess. The amplitude of the D1 component is constrained to lie between 0.5 and 1 times that of the D2 component. The widths of the doublet components are constrained between 0 and 10 {\AA}. We allow the slope of the continuum to lie between $-0.1$ and $0.1$. We impose constraints on the y-intercept of the continuum such that the continuum lies between 0.9 and 1.1 at the wavelength of the D2 component. For additional emission or absorption lines, central wavelengths are constrained to 5 {\AA} below or above the user-specified central wavelength. Amplitudes are constrained between 0 and 2, and widths between 0 and 10 {\AA}. Our $\log{f}$ term is constrained between $-10$ and 1.

For each of the walkers, we discard the first 20 per cent of the generated Markov chain to account for a burn-in time where the walkers settle around the most likely values. Following this, we combine the estimates from each chain together. For each remaining step in the combined chain, we calculate an equivalent width for the Na doublet from the given parameters. We sigma-clip these estimates to ensure any outlying values are ignored, and take the mean of these values as our best fit equivalent width for the doublet, with their standard deviation acting as its associated uncertainty.

In a number of cases, no Na doublet is visible in the spectrum at the expected position. In these cases, the code still attempts to make a fit. After each fit, the spectrum and best fit model are displayed to the user. The user then has the option of either accepting the equivalent width as being associated with a detected absorption line, or marking it as a non-detection. We attempted to implement automatic classification of lines and non-detections based on the strength of the line compared to the noise level but we found that with our small number of spectra, manual vetting gave better results.

For non-detections, we calculate upper limits to the equivalent width of the line by constructing a synthetic line which we would expect to detect. We take the mean and standard deviation of the flux in the normalised spectrum and find the flux which is three standard deviations below the mean flux. We set one minus this value to be the amplitude of our synthetic line. We then estimate the width of the Na line. We do this by noting the instrument used to take the spectrum and checking our fits to the Na line in spectra coming from the same instrument. We take the mean of these widths to be the width of our synthetic line. Using this amplitude and width, we calculate the equivalent width of a Gaussian with amplitude 3$\sigma$ below the mean noise level with a typical width for the Na line, and take this to be our upper limit for the equivalent width from this non-detection.

\begin{figure}
    \centering
    \includegraphics[width = \linewidth]{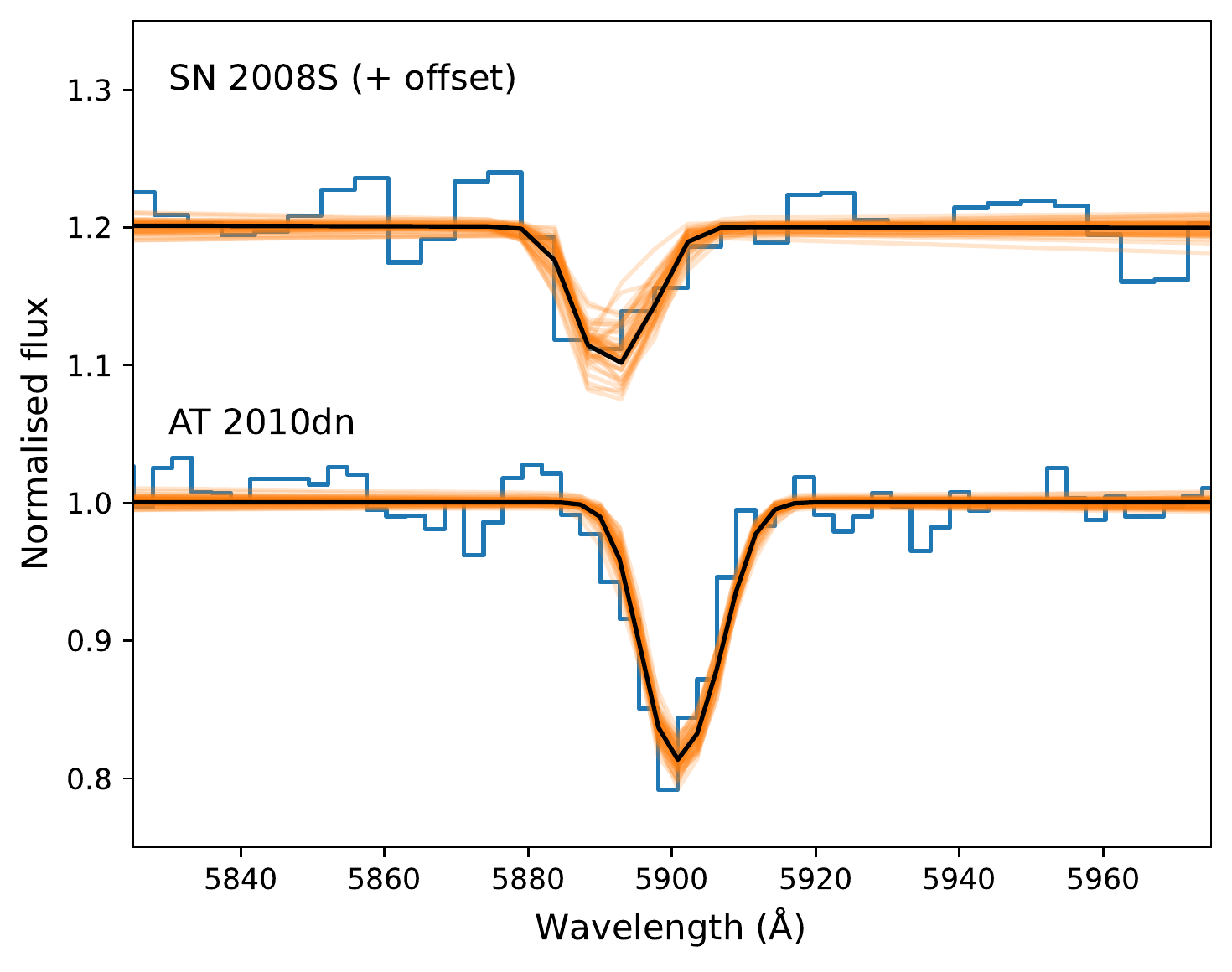}
    \caption{Example of the MCMC fits to two ILRTs: AT 2010dn and SN 2008S at +8.4 days and +46.5 days from explosion, respectively. AT 2010dn shows a strong absorption feature with little variance in the fits produced by the Markov Chain. SN 2008S shows a weaker line compared to the continuum, with a fit showing a larger level of variance. Black line indicates best fit to the doublet, while orange represent a random sample of fits taken from the Markov Chain.}
    \label{fig:spectra}
\end{figure}

Figure \ref{fig:spectra} shows examples of the results of the MCMC code fitting absorption lines in two different spectra. The spectrum of AT 2010dn from +8.4 days shows a strong absorption line clearly distinguishable from the continuum. In contrast, the spectrum of SN 2008S from +46.5 days shows a line which is weak, although still visible in the continuum. Random samples of fits from the Markov chains are displayed in orange, alongside the best fit in black. The fits for AT 2010dn display a low level of spread from the best fit, indicating the low level of uncertainty on this fit due to the strength of the line. Fits for SN 2008S show much more variability from the mean which manifests as a larger uncertainty in the calculated value for equivalent width.

\bsp	
\label{lastpage}
\end{document}